\documentclass[preprint,showpacs,preprintnumbers,amsmath,amssymb,nofootinbib]{revtex4}
\usepackage{graphicx}
\usepackage{dcolumn}
\usepackage{bm}

\def \D {{\cal D}}
\def \be {\begin{equation}}
\def \ee {\end{equation}}
\def \bea{\begin{eqnarray}}
\def \eea{\end{eqnarray}}
\def \no{\nonumber}
\def \bs{{\bf s}}
\def \bq{{\bf q}}
\def \bp{{\bf p}}
\def \bD{{\bf D}}
\def \cU{{\cal U}}
\def \un{{\bf{\hat n}}}
\def \a{\alpha}
\def \b{\beta}
\def \g{\gamma}
\def \z{\zeta}

\begin{document}

\title{Time-Delay Interferometry}

\author{Massimo Tinto}
\email{Massimo.Tinto@jpl.nasa.gov}
\altaffiliation [Also at: ]{LIGO Laboratory, 
California Institute of Technology, Pasadena, CA 91125}
\affiliation{Jet Propulsion Laboratory, California
Institute of Technology, Pasadena, CA 91109}

\author{Sanjeev V. Dhurandhar}
\email{sanjeev@iucaa.ernet.in}
\affiliation{IUCAA, Ganeshkhind, Pune 411 007, India}

\date{\today}

\begin{abstract}
  Equal-arm interferometric detectors of gravitational radiation allow
  phase measurements many orders of magnitude below the intrinsic
  phase stability of the laser injecting light into their arms.  This
  is because the noise in the laser light is common to both arms,
  experiencing exactly the same delay, and thus cancels when it is
  differenced at the photo detector.  In this situation, much lower
  level secondary noises then set overall performance.  If, however,
  the two arms have different lengths (as will necessarily be the case
  with space-borne interferometers), the laser noise experiences
  different delays in the two arms and will hence not directly cancel
  at the detector. In order to solve this problem, a technique
  involving heterodyne interferometry with unequal arm lengths and
  independent phase-difference readouts has been proposed.  It relies
  on properly time-shifting and linearly combining independent Doppler
  measurements, and for this reason it has been called Time-Delay
  Interferometry (or TDI). This article provides an overview of the
  theory and mathematical foundations of TDI as it will be implemented
  by the forthcoming space-based interferometers such as the Laser
  Interferometer Space Antenna (LISA) mission. We have purposely left
  out from this first version of our ``Living Review'' article on TDI
  all the results of more practical and experimental nature, as well
  as all the aspects of TDI that the data analysts will need to
  account for when analyzing the LISA TDI data combinations.  Our
  forthcoming ``second edition'' of this review paper will include
  these topics.
\end{abstract}

\maketitle

\section{Introduction}
\label{Introduction}

Breakthroughs in modern technology have made possible the construction
of extremely large interferometers both on ground and in space for the
detection and observation of gravitational waves~(GW). Several ground
based detectors are being constructed or are already operational
around the globe. These are the LIGO and VIRGO interferometers, which
have arm-lengths of 4 km and 3 km respectively, and the GEO and TAMA
interferometers with arm-lengths of 600 m and 300 m respectively.
These detectors will operate in the high frequency range of GW of
$\sim 1$ Hz to a few kHz. A natural limit occurs on decreasing the
lower frequency cut-off of~$10$ Hz because it is not practical to
increase the arm-lengths on ground and also because of the gravity
gradient noise which is difficult to eliminate below~$10$ Hz. However,
VIRGO and future detectors such as the advanced LIGO, the proposed
LCGT in Japan and the large European detector plan to go to
substantially below $10$ Hz.  Thus, in any case, the ground based
interferometers will not be sensitive below the limiting frequency of
$1$ Hz.  But on the other hand, there exist in the cosmos, interesting
astrophysical GW sources which emit GW below this frequency such as
the galactic binaries, massive and super-massive black-hole binaries
etc. If we wish to observe these sources, we need to go to lower
frequencies. The solution is to build an interferometer in space,
where such noises will be absent and allow the detection of GW in the
low frequency regime.  LISA -~{\it Laser Interferometric Space Antenna
}~- is a proposed mission which will use coherent laser beams
exchanged between three identical spacecraft forming a giant~(almost)
equilateral triangle of side $5 \times 10^6$ kilometers to observe and
detect low frequency cosmic GW.  The ground based detectors and LISA
complement each other in the observation of GW in an essential way,
analogous to the way optical, radio, X-ray, $\gamma$-ray etc.,
observations do for the electromagnetic spectrum. As these detectors
begin to operate, a new era of gravitational astronomy is on the
horizon and a radically different view of the universe is expected to
emerge.

The astrophysical sources that LISA could observe include galactic
binaries, extra-galactic super-massive black-hole binaries and
coalescences, and stochastic GW background from the early universe.
Coalescing binaries are one of the important sources in the LISA
frequency band. These include galactic and extra galactic stellar mass
binaries, and massive and super massive black-hole binaries. The frequency
of the GW emitted by such a system is twice its orbital frequency.
Population synthesis studies indicate a large number of stellar mass
binaries in the frequency range below 2-3 mHz~\cite{PBDH,NYPZ}.  In
the lower frequency range ($ \leq 1$ mHz) there are a large number of
such sources in each of the frequency bins.  Since GW detectors are
omni-directional, it is impossible to resolve an individual source.
These sources effectively form a stochastic GW background referred to
as {\it binary confusion noise}.  

Massive black-hole binaries are interesting both from the
astrophysical and theoretical points of view. Coalescences of massive
blackholes from different galaxies after their merger during growth of
the present galaxies would provide unique new information on galaxy
formation. Coalescence of binaries involving intermediate mass
blackholes could help understand the formation and growth of massive
blackholes. The super massive black hole binaries are strong emitters
of GW and these spectacular events can be detectable beyond red-shift
of $z=1$.  These systems would help to determine the cosmological
parameters independently. And, just as the cosmic microwave background
is left over from the Big Bang, so too should there be a background of
gravitational waves.  Unlike electromagnetic waves, gravitational
waves do not interact with matter after a few Planck times after the
Big Bang, so they do not thermalize.  Their spectrum today, therefore,
is simply a red-shifted version of the spectrum they formed with,
which would throw light on the physical conditions at the epoch of the
early universe.

Interferometric, non-resonant, detectors of gravitational radiation
(with frequency content $0 < f < f_u$) use a coherent train of
electromagnetic waves (of nominal frequency $\nu_0 \gg f_u$) folded
into several beams, and at one or more points where these intersect,
monitor relative fluctuations of frequency or phase (homodyne
detection).  The observed low frequency fluctuations are due to
several causes: (a) frequency variations of the source of the
electromagnetic signal about $\nu_0$, (b) relative motions of the
electromagnetic source and the mirrors (or amplifying transponders)
that do the folding, (c) temporal variations of the index of
refraction along the beams, and, (d) according to general relativity,
to any time-variable gravitational fields present, such as the
transverse-traceless metric curvature of a passing plane gravitational
wave train.  To observe gravitational waves in this way, it is
thus necessary to control, or monitor, the other sources of relative
frequency fluctuations, and, in the data analysis, to use optimal
algorithms based on the different characteristic interferometer
responses to gravitational waves (the signal) and to the other sources
(the noise) \cite{TE95}.  By comparing phases of electromagnetic beams
referenced to the same frequency generator and propagated along
non-parallel equal-length arms, frequency fluctuations of the
frequency reference can be removed and gravitational wave signals at
levels many orders of magnitude lower can be detected.

In the present single-spacecraft Doppler tracking observations, for
instance, many of the noise sources can be either reduced or
calibrated by implementing appropriate microwave frequency links and
by using specialized electronics \cite{Cass}, so the fundamental
limitation is imposed by the frequency (time-keeping) fluctuations
inherent to the reference clock that controls the microwave system.
Hydrogen maser clocks, currently used in Doppler tracking experiments,
achieve their best performance at about $1000$ seconds integration
time, with a fractional frequency stability of a few parts in
$10^{-16}$.  This is the reason why these one-arm interferometers in
space (which have one Doppler readout and a "3-pulse" response to
gravitational waves \cite{EW75}) are most sensitive to millihertz
gravitational waves.  This integration time is also comparable to the
microwave propagation (or "storage") time $2 L/c$ to spacecraft en
route to the outer solar system (for example $L \simeq 5 - 8 \ {\rm
  AU}$ for the Cassini spacecraft) \cite{T02}.

Next-generation low-frequency interferometric gravitational wave
detectors in solar orbits, such as the Laser Interferometer Space
Antenna (LISA) mission \cite{LISA98}, have been proposed to achieve
greater sensitivity to millihertz gravitational waves.  Since the
armlengths of these space-based interferometers can differ by few
percent, the direct recombination of the two beams at a photo detector
will not however effectively remove the laser frequency noise.  This
is because the frequency fluctuations of the laser will be delayed by
different amounts within the two unequal length arms. In order to
cancel the laser frequency noise, the time-varying Doppler data must
be recorded and post-processed to allow for arm-length differences
\cite{TA99}. The data streams will have temporal structure, which can
be described as due to many-pulse responses to $\delta$-function
excitations, depending on time-of-flight delays in the response
functions of the instrumental Doppler noises and in the response to
incident plane-parallel, transverse, and traceless gravitational
waves.

LISA will consists of three spacecraft orbiting the sun.  Each
spacecraft will be equipped with two lasers sending beams to the other
two (${\sim}0.03$ AU away) while simultaneously measuring the beat
frequencies between the local laser and the laser beams received from
the other two spacecraft.  The analysis of TDI presented in this
article will assume a successful prior removal of any first-order
Doppler beat notes due to relative motions \cite{TEA02}, giving six
residual Doppler time series as the raw data of a {\it stationary}
time delay space interferometer.  Following \cite{T98}, \cite{AET99},
\cite{DNV02}, we will regard LISA not as constituting one or more
conventional Michelson interferometers, but rather, in a symmetrical
way, a closed array of six one-arm delay lines between the test
masses.  In this way, during the course of the article, we will show
that it is possible to synthesize new data combinations that cancel
laser frequency noises, and estimate achievable sensitivities of these
combinations in terms of the separate and relatively simple single
arm-responses both to gravitational wave and instrumental noise (cf.
\cite{T98}, \cite{AET99}, \cite{DNV02}).

In contrast to Earth-based interferometers, which operate in the
long-wavelength limit (LWL) (arm lengths $<<$ gravitational wavelength
${\sim}c/f_0$, where $f_0$ is a characteristic frequency of the GW),
LISA will ${\it not}$ operate in the LWL over much of its frequency
band.  When the physical scale of a free mass optical interferometer
intended to detect gravitational waves is comparable to or larger than
the GW wavelength, time delays in the response of the instrument to
the waves, and travel times along beams in the instrument, cannot be
ignored and must be allowed for in computing the detector response
used for data interpretation.  It is convenient to formulate the
instrumental responses in terms of observed differential frequency
shifts $-$ for short, Doppler shifts $-$ rather than in terms of phase
shifts usually used in interferometry, although of course these data,
as functions of time, are interconvertible.

This first review article on TDI is organized as follows. In Section
\ref{SECII} we provide an overview of the physical and historical
motivations of TDI. In Section \ref{SECIII} we summarize the one-arm
Doppler transfer functions of an optical beam between two carefully
shielded test masses inside each spacecraft resulting from (i)
frequency fluctuations of the lasers used in transmission and
reception, (ii) fluctuations due to non-inertial motions of the
spacecraft, (iii) beam-pointing fluctuations and shot noise
\cite{ETA00}.  Among these, the dominant noise is from the frequency
fluctuations of the lasers and is several orders (perhaps 7 or 8)
above the other noises.  This noise must be very precisely removed
from the data in order to achieve the GW sensitivity at the level set
by the remaining Doppler noise sources which are at a much lower level
and which constitute the noise floor after the laser frequency noise
is suppressed. We show that this can be accomplished by shifting and
linearly combining the twelve one-way Doppler data LISA will measure.
The actual procedure can easily be understood in terms of properly
defined time-delay operators that act on the one-way Doppler
measurements. We develop a formalism involving the algebra of the
time-delay operators which is based on the theory of rings and modules
and computational commutative algebra.  We show that the space of all
possible interferometric combinations cancelling the laser frequency
noise is a module over the polynomial ring in which the time-delay
operators play the role of the indeterminates.  The module, in the
literature, is called the Module of Syzygies \cite{DNV02}. We show
that the module can be generated from {\it four} generators, so that
any data combination cancelling the laser frequency noise is simply a
linear combination formed from these generators. We would like to
emphasize that this is the mathematical structure underlying TDI in
LISA.

In Section \ref{SECIV} specific interferometric combinations are then
derived, and their physical interpretations are discussed.  The
expressions for the Sagnac interferometric combinations, ($\alpha,
\beta, \gamma, \zeta$) are first obtained; in particular, the
symmetric Sagnac combination $\zeta$, for which each raw data set
needs to be delayed by only a ${\it single}$ arm transit time,
distinguishes itself against all the other TDI combinations by having
a higher order response to gravitational radiation in the LWL when the
spacecraft separations are equal.  We then express the Unequal-arm
Michelson combinations, ($X, Y, Z$), in terms of the $\alpha$,
$\beta$, $\gamma$, and $\zeta$ combinations with further transit time
delays.  One of these interferometric data combinations would still be
available if the links between one pair of spacecraft were lost.
Other TDI combinations, which rely on only four of the possible six
inter-spacecraft Doppler measurements (denoted $P$, $E$ and $U$) are
also presented. They would of course be quite useful in case of
potential loss of any two inter-spacecraft Doppler measurements.

Time-Delay Interferometry so formulated presumes the
spacecraft-to-spacecraft light-travel-times to be constant in time,
and independent from being up- or down-links.  Reduction of data from
moving interferometric laser arrays in solar orbit will in fact
encounter non-symmetric up- and downlink light time differences that
are significant, and need to be accounted for in order to exactly
cancel the laser frequency fluctuations \cite{S03, CH03, STEA03}. In
Section \ref{SECV} we show that, by introducing a set of non-commuting
time-delay operators, there exists a quite general procedure for
deriving generalized TDI combinations that account for the effects of
time-dependence of the arms. Using this approach it is possible to
derive ``flex-free'' expression for the unequal-arm Michelson
combinations $X_1$, and obtain the generalized expressions for all the
TDI combinations \cite{TEA04}.

In Section \ref{SECVI} we address the question of maximization of the
LISA signal-to-noise-ratio (SNR) to any gravitational wave signal
present in its data.  This is done by treating the SNR as a functional
over the space of all possible TDI combinations.  As a simple
application of the general formula we have derived, we apply our
results to the case of sinusoidal signals randomly polarized and
randomly distributed on the celestial sphere.  We find that the
standard LISA sensitivity figure derived for a single Michelson
Interferometer \cite{ETA00, PTLA02, NPDV03_1} can be improved by a
factor of $\sqrt{2}$ in the low-part of the frequency band, and by
more than $\sqrt{3}$ in the remaining part of the accessible band.
Further, we also show that if the location of the GW source is known,
then as the source appears to move in the LISA reference frame, it is
possible to optimally track the source, by appropriately changing the
data combinations during the course of its trajectory \cite{PTLA02},
\cite{NDPV03_2}. As an example of such type of source, we consider
known binaries within our own galaxy.

This first version of our ``Living Review'' article on TDI does not
include all the results of more practical and experimental nature, as
well as all the aspects of TDI that the data analysts will need to
account for when analyzing the LISA TDI data combinations.  Our
forthcoming ``second edition'' of this review paper will include these
topics. It is worth mentioning that, as of today, the LISA project has
endorsed TDI as its baseline technique for achieving the desired
sensitivity to gravitational radiation.  Several experimental
verifications and tests of TDI are being, and will be, performed at
the NASA and ESA LISA laboratories.  Although significant theoretical
and experimental work has already been done for understanding and
overcoming practical problems related to the implementation of TDI,
more work on both sides of the Atlantic is still needed. Results of
this undergoing effort will be included in the second edition of this
living document.

\section{Physical and historical motivations of TDI}
\label{SECII}

Equal-arm interferometer detectors of gravitational waves can observe
gravitational radiation by cancelling the laser frequency fluctuations
affecting the light injected into their arms. This is done by
comparing phases of split beams propagated along the equal (but
non-parallel) arms of the detector. The laser frequency fluctuations
affecting the two beams experience the same delay within the two
equal-length arms and cancel out at the photodetector where relative
phases are measured. This way gravitational wave signals of
dimensionless amplitude less than $10^{-20}$ can be observed when
using lasers whose frequency stability can be as large as roughly a
few parts in $10^{-13}$.

If the arms of the interferometer have different lengths, however, the
exact cancellation of the laser frequency fluctuations, say $C (t)$,
will no longer take place at the photodetector. In fact, the larger
the difference between the two arms, the larger will be the magnitude
of the laser frequency fluctuations affecting the detector response.
If $L_1$ and $L_2$ are the lengths of the two arms, it is easy to see
that the amount of laser relative frequency fluctuations remaining in
the response is equal to (units in which the speed of light $c = 1$)
\begin{equation}
\Delta C (t) = C(t - 2L_1) - C(t - 2L_2) \ .
\label{DC}
\end{equation}
In the case of a space-based interferometer such as LISA, whose lasers
are expected to display relative frequency fluctuations equal to about
$10^{-13}/\sqrt{Hz}$ in the millihertz band, and whose arms will
differ by a few percent \cite{LISA98}, equation (\ref{DC}) implies the
following expression for the amplitude of the Fourier components of
the uncanceled laser frequency fluctuations (an over imposed tilde
denotes the operation of Fourier transform)
\begin{equation} 
|{\widetilde {\Delta C}} (f)| \simeq |{\widetilde {C}} (f)| \ 
4 \pi f |(L_1 - L_2)| \ .
\label{FDC}
\end{equation}
At $f = 10^{-3}$ Hz, for instance, and assuming $|L_1 - L_2| \simeq
0.5 \ \ {\rm sec}$, the uncanceled fluctuations from the laser are
equal to $6.3 \times 10^{-16}/\sqrt{\rm Hz}$. Since the LISA
sensitivity goal is about $10^{-20}/\sqrt{\rm Hz}$ in this part of the
frequency band, it is clear that an alternative experimental approach
for canceling the laser frequency fluctuations is needed.

\begin{figure}
\centering
\includegraphics[width=7.0 in, angle=0]{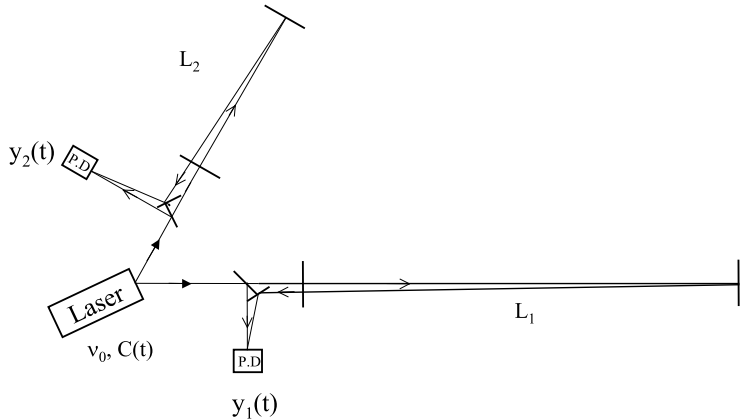}
\caption{Light from a laser is split into two beams, each injected into an
arm formed by pairs of free-falling mirrors. Since the length of the
two arms, $L_1$ and $L_2$, are different, now the light beams from
the two arms are not recombined at one photo detector. Instead each
is separately made to interfere with the light that is injected into
the arms. Two distinct photo detectors are now used, and phase (or
frequency) fluctuations are then monitored and recorded there.}
\label{2unequalarms}
\end{figure}

A first attempt to solve this problem was presented in
\cite{FB84,FBHHV85,FBHHSV89}, and the scheme proposed there can be
understood through Figure \ref{2unequalarms}.  In this idealized model
the two beams exiting the two arms are not made to interfere at a
common photodetector. Rather, each is made to interfere with the
incoming light from the laser at a photodetector, decoupling in this
way the phase fluctuations experienced by the two beams in the two
arms. Now two Doppler measurements are available in digital form, and
the problem now becomes one of identifying an algorithm for digitally
cancelling the laser frequency fluctuations from a resulting new data
combination.

The algorithm they first proposed, and refined subsequently in
\cite{GHTF96}, required processing the two Doppler measurements, say
$y_1 (t)$ and $y_2 (t)$, in the Fourier domain. If we denote with $h_1
(t)$, $h_2 (t)$ the gravitational wave signals entering into the
Doppler data $y_1$, $y_2$ respectively, and with $n_1$, $n_2$ any
other remaining noise affecting $y_1$ and $y_2$ respectively, then the
expressions for the Doppler observables $y_1$, $y_2$ can be written in
the following form
\begin{eqnarray}
y_1(t) & = & C(t - 2L_1) - C(t) + h_1(t) + n_1(t) \ , \ 
\label{2Doppler1}
\\
y_2(t) & = & C(t - 2L_2) - C(t) + h_2(t) + n_2(t) \ .
\label{2Doppler2}
\end{eqnarray}
From Eqs. (\ref{2Doppler1}, \ref{2Doppler2}) it is important to note
the characteristic time signature of the random process $C(t)$ in the
Doppler responses $y_1$ , $y_2$. The time signature of the noise
$C(t)$ in $y_1(t)$, for instance, can be understood by observing that
the frequency of the signal received at time $t$ contains laser
frequency fluctuations transmitted $2L_1$ seconds earlier. By
subtracting from the frequency of the received signal the frequency of
the signal transmitted at time $t$, we also subtract the frequency
fluctuations $C(t)$ with the net result shown in Eq. (\ref{2Doppler1}).

The algorithm for cancelling the laser noise in the Fourier domain
suggested in \cite{FB84} works as follow. If we take an infinitely
long Fourier transform of the data $y_1$, the resulting expression of
$y_1$ in the Fourier domain becomes (see eq. (\ref{2Doppler1}))
\begin{equation}
{\widetilde y}_1 (f) = {\widetilde C} (f) \ [e^{4 \pi i f L_1} - 1] +
{\widetilde h}_1 (f) + {\widetilde n}_1 (f) \ .
\label{2DF1}
\end{equation}
If the arm length $L_1$ is known exactly, we can use the ${\widetilde
  y}_1$ data to estimate the laser frequency fluctuations
${\widetilde C} (f)$. This can be done by dividing ${\widetilde y}_1$
by the transfer function of the laser noise $C$ into the observable
$y_1$ itself. By then further multiplying ${\widetilde y}_1/[e^{4 \pi i f
  L_1} - 1]$ by the transfer function of the laser noise into the
other observable ${\widetilde y}_2$, i.e., $ [e^{4 \pi i f L_2} - 1]$,
and then subtract the resulting expression from ${\widetilde y}_2$ one
accomplishes the cancellation of the laser frequency fluctuations. 

The problem with this procedure is the underlying assumption of being
able to take an infinitely long Fourier transform of the data. Even if
one neglects the variation in time of the LISA arms, by 
taking a finite length Fourier transform of, say, $y_1 (t)$ over
a time interval $T$, the resulting transfer function of the laser noise
$C$ into $y_1$ no longer will be equal to $[e^{4 \pi i f L_1} - 1]$. This
can be seen by writing the expression of the finite length Fourier transform
of $y_1$ in the following way
\begin{equation}
{\widetilde{y}}^{T}_1 \equiv \int_{-T}^{+T} y_1(t) \ e^{2 \pi i f t} \
dt = \int_{-\infty}^{+\infty} y_1(t) \ H(t) \ e^{2 \pi i f t} \ dt \ ,
\label{FiniteF}
\end{equation}
where we have denoted with $H(t)$ the function that is equal to $1$ in
the interval $[-T, +T]$, and zero everywhere else. Equation
(\ref{FiniteF}) implies that the finite-length Fourier transform
${\widetilde{y}}^{T}_1$ of $y_1(t)$ is equal to the convolution in the
Fourier domain of the infinitely long Fourier transform of $y_1 (t)$,
${\widetilde y}_1$, with the Fourier transform of $H(t)$ \cite{JW68}
(i.e. the ``Sinc Function'' of width $1/T$). The key point here is
that we can no longer use the transfer function $[e^{4 \pi i f L_i} -
1] \ , \ i=1, 2$, for estimating the laser noise fluctuations from one
of the measured Doppler data, without retaining residual laser noise
into the combination of the two Doppler data $y_1$, $y_2$ valid in the
case of infinite integration time. The amount of residual laser noise
remaining in the Fourier-based combination described above, as a
function of the integration time $T$ and type of ``window function''
used, was derived in the appendix of \cite{TA99}. There it was shown
that, in order to suppress the residual laser noise below the LISA
sensitivity level identified by secondary noises (such as proof-mass
and optical path noises) with the use of the Fourier-based algorithm
an integration time of about six months was needed.

A solution to this problem was suggested in \cite{TA99}, which works
entirely in the time-domain. From Eqs.(\ref{2Doppler1},
\ref{2Doppler2}) we may notice that, by taking the difference of the
two Doppler data $y_1(t)$, $y_2(t)$, the frequency fluctuations of the
laser now enter into this new data set in the following way
\begin{equation}
y_1(t) - y_2(t) = C(t - 2L_1) - C(t - 2L_2) + h_1(t) - h_2(t) + n_1(t)
- n_2(t) \ .
\label{y1-y2}
\end{equation}
If we now compare how the laser frequency fluctuations enter into Eq.
(\ref{y1-y2}) against how they appear in Eqs. (\ref{2Doppler1},
\ref{2Doppler2}) we can further make the following observation. If we
time-shift the data $y_1(t)$ by the round trip light time in arm $2$,
$y_1(t - 2L_2)$, and subtract from it the data $y_2(t)$ after it has been
time-shifted by the round trip light time in arm $1$, $y_2(t - 2L_1)$, we
obtain the following data set
\begin{eqnarray}
y_1(t - 2L_2) - y_2(t - 2L_1) & = & C(t - 2L_1) - C(t - 2L_2) + h_1(t -
2L_2) - h_2(t - 2L_1) 
\nonumber
\\
&& + n_1(t - 2L_2) - n_2(t - 2L_1) \ .
\label{y1d-y2d}
\end{eqnarray}
In other words, the laser frequency fluctuations enter into $y_1(t) -
y_2(t)$, and $y_1(t - 2L_2) - y_2(t - 2L_1)$ with the same time
structure. This implies that, by subtracting Eq. (\ref{y1d-y2d}) from
Eq. (\ref{y1-y2}) we can generate a new data set that does not contain
the laser frequency fluctuations $C(t)$
\begin{equation}
X \equiv [y_1(t) - y_2(t)] - [y_1(t - 2L_2) - y_2(t - 2L_1)] \ .
\label{XTA99}
\end{equation}
The expression above of the $X$ combination shows that it is possible
to cancel the laser frequency noise in the time domain by properly
time-shifting and linearly combining Doppler measurements recorded by
different Doppler readouts. This in essence is what Time-Delay
Interferometry (TDI) amounts to. In the following sections we will further
elaborate and generalize TDI to the realistic LISA configuration.

\section{Time-Delay Interferometry}
\label{SECIII}

\begin{figure}
\centering
\includegraphics[width=4.0in, angle=0]{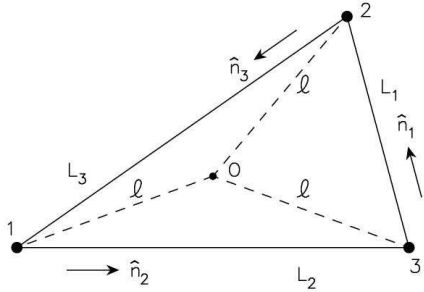}
\caption{Schematic LISA configuration.  The spacecraft are 
  labeled 1, 2, and 3. The optical paths are denoted by $L_i$, $L_i'$
  where the index $i$ corresponds to the opposite spacecraft. The unit
  vectors $\hat{{\bf n}}_i$ point between pairs of spacecraft, with
  the orientation indicated.
\label{Geometry}}
\end{figure}

The description of TDI for LISA is greatly simplified if we adopt the
notation shown in Figure \ref{Geometry} , where the overall geometry
of the LISA detector is defined. There are three spacecraft, six
optical benches, six lasers, six proof-masses and twelve
photodetectors.  There are also six phase difference data going
clock-wise and counter-clockwise around the LISA triangle.  For the
moment we will make the simplifying assumption that the array is
stationary, i.e.  the back and forth optical paths between pairs of
spacecraft are simply equal to their relative distances \cite{S03,
  CH03, STEA03, TEA04}.
\par
Several notations have been used in this context. The double index
notation recently employed in \cite{STEA03} where six quantities are
involved is self-evident. However, when algebraic manipulations are
involved the following notation seems more convenient to use.  The
spacecraft are labeled $1$, $2$, $3$ and their separating distances
are denoted $L_1$, $L_2$, $L_3$, with $L_i$ being opposite spacecraft
$i$. We orient the vertices $1, 2, 3$ clockwise in figure
\ref{Geometry}.  Unit vectors between spacecraft are $\hat n_i$,
oriented as indicated in figure \ref{Geometry}. We index the phase
difference data to be analyzed as follows: The beam arriving at
spacecraft $i$ has subscript $i$ and is primed or unprimed depending
on whether the beam is traveling clockwise or counter-clockwise (the
sense defined here with reference to figure \ref{Geometry}) around the
LISA triangle respectively. Thus, as seen from the figure, $s_{1}$ is
the phase difference time series measured at reception at spacecraft
$1$ with transmission from spacecraft $2$ (along $L_3$).
Similarly, $s_{1}'$ is the phase difference series derived from
reception at spacecraft $1$ with transmission from spacecraft $3$. The
other four one-way phase difference time series from signals exchanged
between the spacecraft are obtained by cyclic permutation of the
indices: $1$ $\to$ $2$ $\to$ $3$ $\to$ $1$.  We also adopt a 
notation for delayed data streams, which will be
convenient later for algebraic manipulations. We define the three
time-delay operators $\D_i, ~i = 1, 2, 3$  where for any data stream
$x(t)$, 
\be 
\D_i x(t) = x(t - L_i) \ ,
\ee 
where $L_i, i = 1, 2, 3$ are the light travel times along the three
arms of the LISA triangle (the speed of light $c$ is assumed to be
unity in this article). Thus, for example, \break $\D_2 s_{1} (t) =
s_{1}(t - L_2)$, $\D_2 \D_3 s_{1}(t) = s_{1}(t - L_2 - L_3) = \D_3
\D_2 s_{1}(t)$, etc.  Note that the operators commute here. This is
because the arm-lengths have been assumed to be constant in time. If
the $L_i$ are functions of time then the operators no longer commute
\cite{CH03,TEA04}, as will be described in section IV. Six more phase 
difference series result
from laser beams exchanged between adjacent optical benches within
each spacecraft; these are similarly indexed as $\tau_{i}, \tau_i', i
= 1, 2, 3$.  The proof-mass-plus-optical-bench assemblies for LISA
spacecraft number $1$ are shown schematically in figure 2.  The photo
receivers that generate the data $s_{1}$, $s_{1}'$, $\tau_{1}$, and
$\tau_{1}'$ at spacecraft $1$ are shown.  The phase fluctuations from
the six lasers, which need to be canceled, can be represented by six
random processes $p_{i}, p_{i}'$, where $p_{i}, p_{i}'$ are the phases
of the lasers in spacecraft $i$ on the left and right optical benches
respectively as shown in the figure. Note that this notation is in the
same spirit as in the references \cite{TEA02, STEA03} in which moving
spacecraft arrays have been analyzed.
\par
We extend the cyclic terminology so that at vertex $i$ ($i = 1, 2, 3$)
the random displacement vectors of the two proof masses are
respectively denoted by $\vec \delta_{i}(t), \vec \delta_{i}'(t)$, and
the random displacements (perhaps several orders of magnitude greater)
of their optical benches are correspondingly denoted by $\vec
\Delta_{i}(t), \vec \Delta_{i}'(t)$ where the primed and unprimed
indices correspond to the right and left optical benches respectively.
As pointed out in \cite{ETA00}, the analysis does \underline {not}
assume that pairs of optical benches are rigidly connected, i.e. $\vec
\Delta_{i} \neq \vec \Delta_{i}'$, in general.  The present LISA
design shows optical fibers transmitting signals both ways between
adjacent benches.  We ignore time-delay effects for these signals and
will simply denote by $\mu_i(t)$ the phase fluctuations upon
transmission through the fibers of the laser beams with frequencies
$\nu_{i}$, and $\nu_{i}'$.  The $\mu_i (t)$ phase shifts within a
given spacecraft might not be the same for large frequency differences
$\nu_{i} - \nu_{i}'$. For the envisioned frequency differences (a few
hundred megahertz), however, the remaining fluctuations due to the
optical fiber can be neglected \cite{ETA00}.  It is also assumed that
the phase noise added by the fibers is independent of the direction of
light propagation through them.  For ease of presentation, in what
follows we will assume the center frequencies of the lasers to be the
same, and denote this frequency by $\nu_0$.

The laser phase noise in $s_{3}'$ is therefore equal to $\D_1 p_{2}(t)
- p_{3}'(t)$.  Similarly, since $s_{2}$ is the phase shift measured on
arrival at spacecraft $2$ along arm $1$ of a signal transmitted from
spacecraft $3$, the laser phase noises enter into it with the
following time signature: $\D_1 p_{3}'(t) - p_{2}(t)$.  Figure 2
endeavors to make the detailed light paths for these observations
clear.  An outgoing light beam transmitted to a distant spacecraft is
routed from the laser on the local optical bench using mirrors and
beam splitters; this beam does not interact with the local proof mass.
Conversely, an {\it {incoming}} light beam from a distant spacecraft
is bounced off the local proof mass before being reflected onto the
photo receiver where it is mixed with light from the laser on that
same optical bench. The inter-spacecraft phase data are denoted
$s_{1}$ and $s_{1}'$ in figure 2.
\begin{figure}
\centering
\includegraphics[width=5.0 in, angle=0]{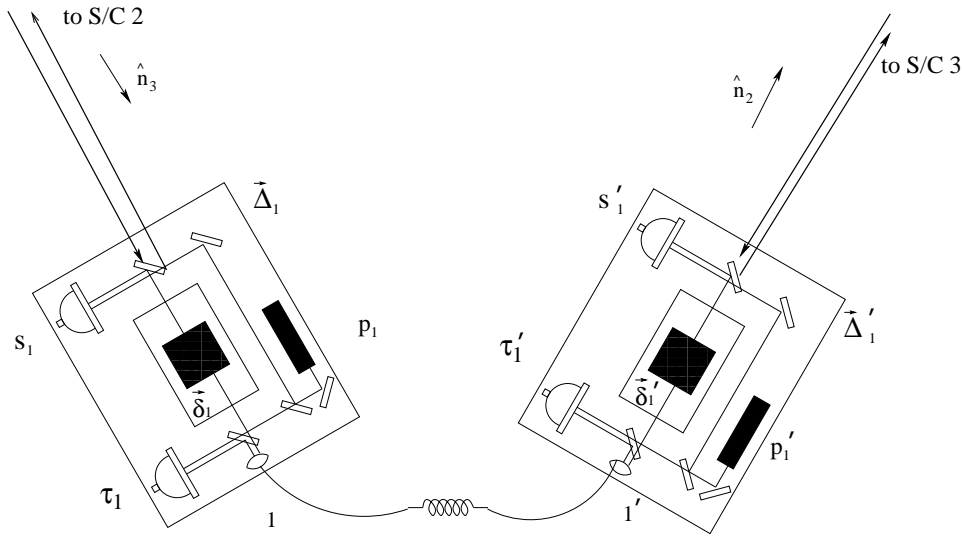}
\caption{Schematic diagram of proof-masses-plus-optical-benches
  for a LISA spacecraft.  The left-hand bench reads out the phase
  signals $s_{1}$ and $\tau_{1}$.  The right hand bench analogously
  reads out $s_{1}'$ and $\tau_{1}'$.  The random displacements of the
  two proof masses and two optical benches are indicated (lower case
  $\vec \delta_{i}, \vec \delta_{i}'$ for the proof masses, upper case
  $\vec \Delta_{i}, \Delta_{i}'$ for the optical benches).}
\end{figure}
Beams between adjacent optical benches within a single spacecraft are
bounced off proof masses in the opposite way.  Light to be {\it
  {transmitted}} from the laser on an optical bench is {\it {first}}
bounced off the proof mass it encloses and then directed to the other
optical bench.  Upon reception it does {\it not} interact with the
proof mass there, but is directly mixed with local laser light, and
again down converted. These data are denoted $\tau_{1}$ and
$\tau_{1}'$ in figure 2.

The expressions for the $s_{i}, s_{i}'$ and $\tau_{i}, \tau_{i}'$
phase measurements can now be developed from figures 1 and 2, and they
are for the particular LISA configuration in which all the lasers have
the same nominal frequency $\nu_0$, and the spacecraft are stationary
with respect to each other. Consider the $s_{1}' (t)$ process
(equation (\ref{eq:tre}) below).  The photo receiver on the right bench
of spacecraft $1$, which (in the spacecraft frame) experiences a
time-varying displacement $\vec \Delta_{1}'$, measures the phase
difference $s_{1}'$ by first mixing the beam from the distant optical
bench $3$ in direction $\hat n_2$, and laser phase noise $p_{3}$ and
optical bench motion $\vec \Delta_{3}$ that have been delayed by
propagation along $L_2$, after one bounce off the proof mass ($\vec
\delta_{1}'$), with the local laser light (with phase noise $p_{1}'$).
Since for this simplified configuration no frequency offsets are
present, there is of course no need for any heterodyne conversion
\cite{TEA02}.

In equation (\ref{eq:due}) the $\tau_{1}$ measurement results from light
originating at the right-bench laser ($p_{1}'$, $\vec \Delta_{1}'$),
bounced once off the right proof mass ($\vec \delta_{1}'$), and
directed through the fiber (incurring phase shift $\mu_1(t)$), to the
left bench, where it is mixed with laser light ($p_{1}$).  Similarly
the right bench records the phase differences $s_{1}'$ and
$\tau_{1}'$.  The laser noises, the gravitational wave signals, the
optical path noises, and proof-mass and bench noises, enter into the
four data streams recorded at vertex $1$ according to the following
expressions \cite{ETA00}:

\begin{eqnarray}
s_{1} & = & s^{\rm gw}_{1} + s^{\rm opt. \ path}_{1} + 
\D_3 p_{2}' - p_{1} + \nu_{0} \ \left[\ - 2 {\hat n_3} \cdot 
{\vec \delta_{1}} + {\hat n_3} \cdot {\vec \Delta_{1}} +  
{\hat n_3} \cdot \D_3 {\vec \Delta_{2}'} \right] \ ,
\label{eq:uno}
\\
\tau_{1} & = & p_{1}' - p_{1} - 
 \ 2 \ \nu_{0} \ {\hat n_2} \cdot ({\vec \delta_{1}'} - {\vec
  \Delta_{1}'}) + \mu_1 \ .
\label{eq:due}
\\
s_{1}' & = &  s^{'{\rm gw}}_{1} + s^{'{\rm opt. \ path}}_{1} +
\D_2 p_{3} - p_{1}' + \nu_{0} \ \left[ 2 {\hat n_2} \cdot {\vec \delta_{1}'}  -   
{\hat n_2} \cdot {\vec \Delta_{1}'} - {\hat n_2} \cdot \D_2 {\vec
  \Delta_{3}} \right] \ ,
\label{eq:tre}
\\
\tau_{1}' & = &  p_{1} - p_{1}' + 2 \ \nu_{0} \ {\hat n_3} \cdot 
({\vec \delta_{1}} - {\vec \Delta_{1}}) + \mu_1 \ .
\label{eq:quattro}
\end{eqnarray}
\noindent
Eight other relations, for the readouts at vertices 2 and 3, are given
by cyclic permutation of the indices in equations
(\ref{eq:uno})-(\ref{eq:quattro}).

The gravitational wave phase signal components, $s^{\rm gw}_{i},
s^{'{\rm gw}}_{i} \ , \ i = 1, 2, 3$, in equations (\ref{eq:uno}) and
(\ref{eq:tre}) are given by integrating with respect to time the
equations (1), and (2) of reference \cite{AET99}, which relate metric
perturbations to optical frequency shifts. The optical path phase
noise contributions, $s^{\rm opt. \ path}_{i}, s^{'{\rm opt. \ 
    path}}_{i}$, which include shot noise from the low signal-to-noise
ratio (SNR) in the links between the distant spacecraft, can be
derived from the corresponding term given in \cite{ETA00}. The
$\tau_{i}, \tau_{i}'$ measurements will be made with high SNR so that
for them the shot noise is negligible.

\section{Albegraic approach to cancelling laser and optical bench
  noises}
\label{SECIV}

In ground based detectors the arms are chosen to be of equal length so
that the laser light experiences identical delay in each arm of the
interferometer.  This arrangement precisely cancels the laser
frequency/phase noise at the photodetector. The required sensitivity
of the instrument can thus only be achieved by near exact cancellation
of the laser frequency noise. However, in LISA it is impossible to
achieve equal distances between spacecraft and the laser noise cannot
be cancelled in this way. It is possible to combine the recorded data
linearly with suitable time-delays corresponding to the three
arm-lengths of the giant triangular interferometer so that the laser
phase noise is cancelled. Here we present a {\it systematic method}
based on modules over polynomial rings which guarantees all the data
combinations that cancel both the laser phase and the optical
bench motion noises.
\par
We first consider the simpler case, where we ignore the optical-bench
motion noise and consider only the laser phase noise. We do this
because the algebra is somewhat simpler and the method is easy to
apply.  The simplification amounts to physically considering each
spacecraft rigidly carrying the assembly of lasers, beam-splitters and
photodetectors. The two lasers on each spacecraft could be considered
to be locked, so effectively there would be only one laser on each
spacecraft. This mathematically amounts to setting $\vec \Delta_i =
\vec \Delta_i'= 0$ and $p_i = p_i'$.  The scheme we describe here for
laser phase noise can be extended in a straight forward way to include
optical bench motion noise, which we address in the last part of this
section.
\par 
The data combinations, when only the laser phase noise is considered,
consist of the six suitably delayed data streams (inter-spacecraft),
the delays being integer multiples of the light travel times between
spacecraft, which can be conveniently expressed in terms of
polynomials in the three delay operators $\D_1, \D_2, \D_3$. The laser
noise cancellation condition puts three constraints on the six
polynomials of the delay operators corresponding to the six data
streams. The problem therefore consists of finding six tuples of
polynomials which satisfy the laser noise cancellation constraints.
These polynomial tuples form a module \footnote{A module is an abelian
  group over a {\it ring} as contrasted with a vector space which is
  an abelian group over a field. The scalars form a ring and just like
  in a vector space, scalar multiplication is defined. However, in a
  ring the multiplicative inverses do not exist in general for the
  elements, which makes all the difference!} called in the literature,
the {\it module of syzygies}. There exist standard methods for
obtaining the module, by that we mean, methods for obtaining the
generators of the module so that the linear combinations of the
generators generate the entire module. The procedure first consists of
obtaining a Gr\"obner basis for the ideal generated by the
coefficients appearing in the constraints. This ideal is in the
polynomial ring in the variables $\D_1, \D_2, \D_3$ over the domain of
rational numbers~(or integers if one gets rid of the denominators). To
obtain the Gr\"obner basis for the ideal, one may use the Buchberger
algorithm or use an application such as Mathematica \cite{Wolf02}.
From the Gr\"obner basis there is a standard way to obtain a
generating set for the required module. This procedure has been
described in the literature~\cite{becker,KR}. We thus obtain seven
generators for the module.  However, the method does not guarantee a
minimal set and we find that a generating set of 4 polynomial six
tuples suffice to generate the required module.  Alternatively, we can
obtain generating sets by using the software Macaulay 2.
\par
The importance of obtaining more data combinations is evident: they
provide the necessary redundancy - different data combinations produce
different transfer functions for GW and the system noises so
specific data combinations could be optimal for given astrophysical
source parameters in the context of maximizing SNR, detection
probability, improving parameter estimates etc.

\subsection{Cancellation of laser phase noise}
 
We now only have six data streams: $s_i$ and $s_i'$ where, $i =
1,2,3$. These can be regarded as 3 component vectors $\bs$ and $\bs'$
respectively. The six data streams with terms containing {\it only}
the laser frequency noise are: 
\bea
s_1 &=& \D_3 p_2 - p_1 , \no \\
s_1' &=& \D_2 p_3 - p_1 ,
\label{beams} 
\eea  
\noindent
and their cyclic permutations.

Note that we have excluded intentionally from the data, additional
phase fluctuations due to the GW signal, and noises such as the
optical-path noise, proof-mass noise etc. Since our immediate goal is
to cancel the laser frequency noise we have only kept the relevant
terms. Combining the streams for cancelling the laser frequency noise
will introduce transfer functions for the other noises and the GW
signal. This is important and will be discussed subsequently in the
article.

{\it The goal of the analysis is to add suitably delayed beams
  together so that the laser frequency noise terms add up to zero.}
This amounts to seeking data combinations that cancel the laser
frequency noise. In the notation/formalism that we have invoked, the
delay is obtained by applying the operators $\D_k$ to the beams $s_i$
and $s_i'$. A delay of $k_1 L_1 + k_2 L_2 + k_3 L_3$ is represented by
the operator $\D_1^{k_1} \D_2^{k_2} \D_3^{k_3}$ acting on the data,
where $k_1, k_2$ and $k_3$ are integers. In general a polynomial in
$\D_k$, which is a polynomial in three variables, applied to say $s_1$
combines the same data stream $s_1(t)$ with different time-delays of
the form $k_1 L_1 + k_2 L_2 + k_3 L_3$. This notation conveniently
rephrases the problem. One must find six polynomials say $q_i (\D_1,
\D_2, \D_3), \ q_i' (\D_1, \D_2, \D_3), \ i=1,2,3 $ such that: 
\be
\sum_{i = 1}^3 q_i s_i + q_i' s_i' = 0 \ .
\label{cncl}
\ee 
The zero on the R.H.S. of the above equation signifies zero laser
phase noise.
\par
It is useful to express Eq.~(\ref{beams}) in matrix form. This allows 
us to obtain a matrix operator equation whose solutions are $\bq$ and $\bq'$ 
where $q_i$ and $q_i'$ are written as column vectors. We can 
similarly express $s_i, s_i', p_i$ as  column vectors $\bs, \bs', \bp$ 
respectively. In matrix form Eq.~(\ref{beams}) become:
\be 
\bs = \bD^T \ \cdot  \bp \ , ~~~~~~~ \bs' =  \bD \ \cdot  \bp \ ,
\ee
where, $\bD$ is a $3 \times 3$ matrix given by,
\be
\bD = \left(\begin{array}{ccc}
  -1 & 0 & \D_2 \\
  \D_3 & -1 & 0 \\
  0 & \D_1 & -1
\end{array}\right) \,.
\label{Dmat}
\ee
The exponent `$T$' represents the transpose of the matrix. 
Eq.~(\ref{cncl}) becomes:
\be
\bq^T \cdot \bs + \bq'^T \cdot \bs' = (\bq^T \cdot \bD^T  + \bq'^T \cdot \bD ) \cdot \bp = 0  \, .
\ee
where we have taken care to put $\bp$ on the right-hand-side of the
operators. Since the above equation must be satisfied for an arbitrary
vector $\bp$, we obtain a matrix equation for the polynomials $(\bq,
\bq')$:
\be
\bq^T \cdot \bD^T + \bq' \cdot \bD = 0 \, .
\label{opeq}
\ee 
Note that since the $\D_k$ commute, the order in writing these
operators is unimportant. In mathematical terms, the polynomials form
a commutative ring.

\subsection{Cancellation of laser phase noise in the unequal arm
  interferometer}

The use of commutative algebra is very conveniently illustrated with
the help of the simpler example of the unequal arm interferometer.
Here there are only two arms instead of three as we have for LISA and
the mathematics is much simpler and so it easy to see both physically
and mathematically how commutative algebra can be applied to this
problem of laser phase noise cancellation. The procedure is well known
for the unequal arm interferometer, but here we will describe the same
method but in terms of the delay opertors that we have introduced.
\par
\begin{figure}
\centering
\includegraphics[width=3in, angle=0]{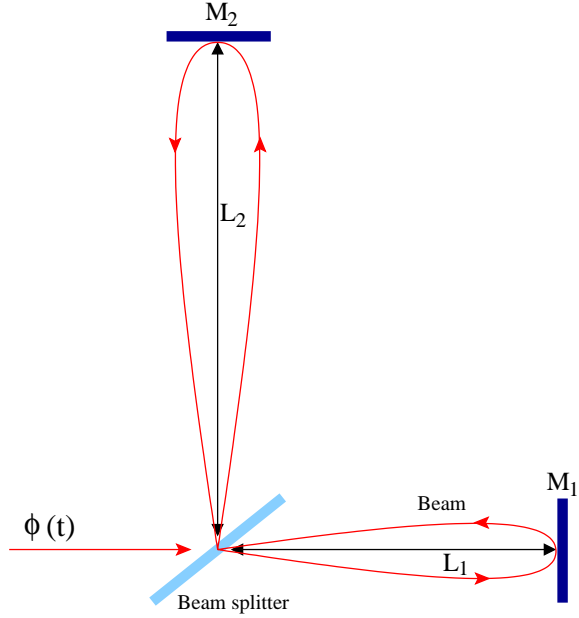}
\caption{Schematic diagram of the unequal-arm Michelson
  interferometer. The beam shown corresponds to the term $(\D_1^2 -
  1)(\D_2^2 - 1) \phi_ (t)$ in $X(t)$ which is first sent around arm 1
  followed by arm 2. The second beam (not shown) is first sent around
  arm 2 and then through arm 1.  The difference in these two beams
  constitutes $X(t)$.}
\label{XU}
\end{figure}

Let $\phi (t)$ denote the laser phase noise entering the laser cavity
as shown in Fig. \ref{XU}. Consider this light $\phi (t)$ making a
round trip around arm 1 whose length we take to be $L_1$. If we
interfere this phase with the incoming light we get the phase $\phi_1
(t)$, where, 
\be 
\phi_1 (t) = \phi (t - 2 L_1) - \phi (t) \equiv (\D_1^2 - 1) \phi (t) .  
\ee 
The second expression we have written in
terms of the delay operators. This makes the procedure transparent as
we shall see.  We can do the same for the arm 2 to get another phase
$\phi_2 (t)$, where, 
\be 
\phi_2 (t) = \phi (t - 2 L_2) - \phi (t) \equiv (\D_2^2 - 1) \phi (t) .  
\ee 
Clearly, if $L_1 \neq L_2$, then
the difference in phase $\phi_2 (t) - \phi_1 (t)$ is not zero and the
laser phase noise does not cancel out. However, if one further delays
the phases $\phi_1 (t)$ and $\phi_2 (t)$ and constructs the following
combination: 
\be 
X(t) = [\phi_2 (t - 2 L_1) - \phi_2 (t)] - [\phi_1 (t - 2 L_2) - \phi_1 (t)] , 
\label{X} 
\ee 
then the laser phase noise does cancel out. We have already encountered this combination at the 
end of section II, Eq.(\ref{XTA99}), when it was first proposed by Tinto and Armstrong 
in reference \cite{TA99}.  
The cancellation of laser frequency noise becomes obvious from the operator
algebra in the following way. In the operator notation: 
\bea
X(t) &=& (\D_1^2 - 1) \phi_2 (t) - (\D_2^2 - 1) \phi_1 (t) \, \no \\
&=& [(\D_1^2 - 1)(\D_2^2 - 1) - (\D_2^2 - 1)(\D_1^2 - 1)] \phi (t) \, \no \\
&=& 0 .  
\eea
 From this one immediately sees that just the
commutativity of the operators has been used to cancel the laser phase
noise. The basic idea was to compute the lowest common multiple
(L.C.M.) of the polynomials $\D_1^2 - 1$ and $\D_2^2 - 1$ (in this
case the L.C.M. is just the product, because the polynomials are
relatively prime) and use this fact to construct $X(t)$ in which the
laser phase noise is cancelled. The operation is shown physically in
Fig. \ref{XU}.
The notions of commutativity of polynomials, L.C.M. etc. belong to the
field of commutative algebra. In fact we will be using the notion of a
Gr\"obner basis which is in a sense the generalization of the notion
of the greatest common divisor or in short gcd.  Since LISA has three
spacecraft and six inter-spacecraft beams, the problem of the unequal
arm interferometer only gets technically more complex; in principle
the problem is the same as in this simpler case. Thus the simple
operations which were performed here to obtain a laser noise free
combination $X(t)$ are not sufficient and more sophisticated methods
need to be adopted from the field of commutative algebra. We address
this problem in the forthcoming text.

\subsection{The Module of Syzygies}

Eq.~(\ref{opeq}) has non-trivial solutions. Several solutions have
been exhibited in~\cite{AET99,ETA00}. We merely mention these
solutions here; in the next section we will discuss them in detail.
The solution $\zeta$ is given by $-\bq^T = \bq'^T = (\D_1, \D_2,
\D_3)$. The solution $\alpha$ is described by $\bq^T = -(1, \D_3, \D_1
\D_3)$ and $\bq'^T = (1, \D_1 \D_2, \D_2)$.  The solutions $\beta$ and
$\gamma$ are obtained from $\alpha$ by cyclically permuting the
indices of $\D_k, \bq$ and $\bq'$. These solutions are important,
because they consist of polynomials with lowest possible degrees and
thus are simple. Other solutions containing higher degree polynomials
can be generated conveniently from these solutions. Since the system
of equations is linear, linear combinations of these solutions are
also solutions to Eq.~(\ref{opeq}).
\par
However, it is important to realize that we do not have a vector space
here. Three independent constraints on a six tuple do not produce a
space which is necessarily generated by three basis elements. This
conclusion would follow if the solutions formed a vector space but
they do not. The polynomial six-tuple $\bq, \bq'$ can be multiplied by
polynomials in $\D_1, \D_2, \D_3$ (scalars) which do not form a field.
So that the inverse in general does not exist within the ring of
polynomials. We therefore have a module over the ring of polynomials
in the three variables $\D_1, \D_2, \D_3$.  First we present the
general methodology for obtaining the solutions to~(\ref{opeq}) and
then apply it to equations~(\ref{opeq}).
\par
There are three linear constraints on the polynomials given by the
equations~(\ref{opeq}). Since the equations are linear the solutions
space is a submodule of the module of six-tuples of polynomials. The
module of six-tuples is a free module, i.e. it has six basis elements
that not only generate the module but are linearly independent. A
natural choice of the basis is $f_m = (0, ..., 1, ...,0)$ with 1 in
the $m$-th place and 0 everywhere else; $m$ runs from 1 to 6. The
definitions of generation~(spanning) and linear independence are the
same as that for vector spaces. A free module is essentially like a
vector space. But our interest lies in its submodule which need not be
free and need not have just three generators as it would seem if we
were dealing with vector spaces.
\par
The problem at hand is of finding the generators of this submodule
i.e. any element of the submodule should be expressible as a linear
combination of the generating set. In this way the generators are
capable of spanning the full submodule or generating the submodule. In
order to achieve our goal, we rewrite the Eq.~(\ref{opeq}) explicitly
component-wise:
\bea
q_1 + q_1' - \D_3 q_2' - \D_2 q_3 &=& 0 \ , \no \\
q_2 + q_2' - \D_1 q_3' - \D_3 q_1 &=& 0 \ , \no \\
q_3 + q_3' - \D_2 q_1' - \D_1 q_2 &=& 0 \ .
\label{lneq}
\eea 

The first step is to use Gaussian elimination to obtain $q_1$ and
$q_2$ in terms of $q_3, q_1', q_2', q_3'$:

\bea
q_1 &=& - q_1' + \D_3 q_2' + \D_2 q_3 \ , \no \\
q_2 &=& - q_2' + \D_1 q_3' + \D_3 q_1 \no \\
&=& - \D_3 q_1' - (1 - \D_3^2) q_2' + \D_1 q_3' + \D_2 \D_3 q_3 \ ,
\label{gauss}
\eea
and then substitute these values in the third equation to obtain a
linear implicit relation between $q_3, q_1', q_2', q_3'$. We then
have:
\be
(1 - \D_1 \D_2 \D_3) q_3 + (\D_1 \D_3 - \D_2) q_1' + \D_1 (1 - \D_3^2) q_2' + (1 - \D_1^2) q_3' = 0 \ . 
\label{implct}
\ee 
Obtaining solutions to Eq.~(\ref{implct}) amounts to solving the
problem since the the remaining polynomials $q_1, q_2$ have been
expressed in terms of $q_3, q_1', q_2', q_3'$ in~(\ref{gauss}). Note
that we cannot carry on the Gaussian elimination process any further,
because none of the polynomial coefficients appearing in
Eq.(\ref{implct}) have an inverse in the ring.
\par
We will assume that the polynomials have rational coefficients {\it
  i.e} the coefficients belong to ${\cal Q}$ the field of the rational
numbers.  The set of polynomials form a ring - the polynomial ring in
three variables which we denote by ${\cal R}={\cal Q}[\D_1, \D_2,
\D_3]$.  The polynomial vector $(q_3, q_1', q_2', q_3')\, \in \, {\cal
  R}^4$. The set of solutions to~(\ref{implct}) is just the {\it
  kernel} of the homomorphism $\varphi:\, {\cal R}^4 \rightarrow {\cal
  R}$ where the polynomial vector $(q_3, q_1', q_2', q_3')$ is mapped
to the polynomial \break $(1 - \D_1 \D_2 \D_3) q_3 + (\D_1 \D_3 -
\D_2) q_1' + \D_1 (1 - \D_3^2) q_2' + (1 - \D_1^2) q_3' $. Thus the
solution space $ ker \varphi$ is a submodule of ${\cal R}^4$. It is
called the {\it module of syzygies} in the literature.  The generators
of this module can be obtained from standard methods available in the
literature. We briefly outline the method given in the books by Becker
et al.~\cite{becker} and Kreuzer and Robbiano~\cite{KR} below.  The
details have been included in appendix A.

\subsection{Gr\"obner Basis}

The first step is to obtain the Gr\"obner basis for the ideal $\cU$ generated 
by the coefficients in Eq.~(\ref{implct}):
\be 
u_1 = 1 - \D_1 \D_2 \D_3,~~u_2 = \D_1 \D_3 - \D_2,~~u_3 =
\D_1 (1 - \D_3^2),~~u_4 = 1 - \D_1^2 \ . 
\label{idgen}
\ee
The ideal $\cU$ consists of linear combinations of the form $\sum v_i
u_i$ where $v_i$, $i = 1, ..., 4$ are polynomials in the ring ${\cal
  R}$.  There can be several sets of generators for $\cU$.  A
Gr\"obner basis is a set of generators which is `small' in a specific
sense.
\par
There are several ways to look at the theory of Gr\"obner basis. One
way is, suppose we are given polynomials $g_1, g_2, ..., g_m$ in one
variable over say ${\cal Q}$ and we would like to know whether another
polynomial $f$ belongs to the ideal generated by the $g$'s. A good way
to decide the issue would be to first compute the gcd~(greatest common
divisor) $g$ of $g_1, g_2, ..., g_m$ and check whether $f$ is a
multiple of $g$. One can achieve this by doing the long division of
$f$ by $g$ and checking whether the remainder is zero. All this is
possible because ${\cal Q}[x]$ is a Euclidean domain and also a
principle ideal domain~(PID) wherein any ideal is generated by a
single element. Therefore we have essentially just one polynomial -
the gcd - which generates the ideal generated by $g_1, g_2, \dots ,
g_m$. The ring of integers or the ring of polynomials in one variable
over any field are examples of PIDs whose ideals are generated by
single elements. However, when we consider more general rings~(not
PIDs) like the one we are dealing with here, we do not have a single
gcd but a set of several polynomials which generates an ideal in
general. A Gr\"obner basis of an ideal can be thought of as a
generalization of the gcd. In the univariate case, the Gr\"obner basis
reduces to the gcd.
\par
Gr\"obner basis theory generalizes these ideas to multivariate
polynomials which are neither Euclidean rings nor PIDs. Since there is
in general not a single generator for an ideal, Gr\"obner basis theory
comes up with the idea of dividing a polynomial with a {\it set} of
polynomials, the set of generators of the ideal, so that by successive
divisions by the polynomials in this generating set of the given
polynomial, the remainder becomes zero.  Clearly, every generating set
of polynomials need not possess this property.  Those special
generating sets that do possess this property~(and they exist!)  are
called Gr\"obner bases. In order for a division to be carried out in a
sensible manner, an order must be put on the ring of polynomials, so
that the final remainder after every division is strictly smaller than
each of the divisors in the generating set.  A natural order exists on
the ring of integers or on the polynomial ring ${\cal Q}(x)$; the
degree of the polynomial decides the order in ${\cal Q}(x)$. However,
even for polynomials in two variables there is no natural order
apriori (Is $x^2 + y$ greater or smaller than $x + y^2$?). But one
can, by hand as it were, put an order on such a ring by saying $x >>
y$, where $ >> $ is an order, called the lexicographical order. We
follow this type of order, $\D_1 >> \D_2 >> \D_3$ and ordering
polynomials by considering their highest degree terms. It is possible
to put different orderings on a given ring which then produce
different Gr\"obner bases. Clearly, a Gr\"obner basis must have
`small' elements so that division is possible and every element of the
ideal when divided by the Gr\"obner basis elements leaves zero
remainder, {\it i.e.}  every element modulo the Gr\"obner basis
reduces to zero.
\par
In the literature, there exists a well-known algorithm called the the
Buchberger algorithm which may be used to obtain the Gr\"obner basis
for a given set of polynomials in the ring. So a Gr\"obner basis of
$\cU$ can be obtained from the generators $u_i$ given in
Eq.~(\ref{idgen}) using this algorithm. It is essentially again a
generalization of the usual long division that we perform on
univariate polynomials.  More conveniently, we prefer to use the
well known application `Mathematica'. Mathematica yields a 3 element
Gr\"obner basis ${\cal G}$ for $\cU$: 
\be {\cal G} = \{\D_3^2 - 1,
\D_2^2 - 1, \D_1 - \D_2 \D_3 \} \, .  
\ee 
One can easily check that
all the $u_i$ of Eq.~(\ref{idgen}) are linear combinations of the
polynomials in ${\cal G}$ and hence ${\cal G}$ generates $\cU$. One
also observes that the elements look `small' in the order mentioned
above. However, one can satisfy oneself that ${\cal G}$ is a Gr\"obner
basis by using the standard methods available in the literature.  One
method consists of computing the S-polynomials~(see Appendix A) for
all the pairs of the Gr\"obner basis elements and checking whether
these reduce to zero modulo ${\cal G}$.

This Gr\"obner basis of the ideal $\cU$ is then used to obtain the
generators for the module of syzygies. Note that although the
Gr\"obner basis depends on the order we choose among the $\D_k$, the
module itself is {\it independent} of the order \cite{becker}.

\subsection{Generating Set for the Module of Syzygies}
\label{sec:genset} 

The generating set for the module is obtained by further following the 
procedure in the literature~\cite{becker,KR}. The details are given 
in Appendix A, specifically for our case.  We obtain 7 generators for 
the module. These generators do not form a minimal set and there are 
relations between them; in fact this method does not guarantee a minimum 
set of generators. These generators can be expressed as linear combinations 
of $\alpha, \beta, \gamma, \zeta$ and also in terms of $X^{(1)}, X^{(2)}, 
X^{(3)}, X^{(4)}$ given below in Eq.~(\ref{gen6}). The importance  in 
obtaining the 7 generators is that the standard theorems guarantee that 
these 7 generators do in fact generate the required module. Therefore, from 
this proven set of generators we can check whether a particular set is 
in fact a generating set. We present several generating sets below. 
\par
Alternatively, we may use a software package called {\it Macaulay 2} which directly 
calculates the generators given the the equations~(\ref{lneq}). Using 
{\it Macaulay 2}, we obtain six  generators. Again, Macaulay's algorithm 
does not yield a minimal set; we can express the last two generators in 
terms of the first four. Below we list this  smaller  set of 
four generators in the order $X = (q_1, q_2, q_3, q_1', q_2', q_3')$:
\bea
X^{(1)} &=& (\D_2 - \D_1 \D_3, 0, 1 - \D_3^2, 0, \D_2 \D_3 - \D_1, \D_3^2 - 1) \ , \no \\
X^{(2)} &=& (- \D_1, - \D_2, - \D_3, \D_1, \D_2, \D_3) \ , \no \\
X^{(3)} &=& (-1, - \D_3, - \D_1 \D_3, 1, \D_1 \D_2, \D_2) \ , \no \\
X^{(4)} &=& (- \D_1 \D_2, -1, - \D_1, \D_3, 1, \D_2 \D_3) \ . 
\label{gen6}
\eea
Note that the last three generators are just $X^{(2)} = \zeta, X^{(3)} =
\alpha,  X^{(4)} = \beta$. An extra generator $X^{(1)}$ is needed 
to generate all  the solutions.
\par
Another set of generators which may be useful for further work is a
Gr\"obner basis of a module. The concept of a Gr\"obner basis of an
ideal can be extended to that of a Gr\"obner basis of a submodule of
$(K[x_1, x_2, ..., x_n])^m$ where $K$ is a field, since a module over
the polynomial ring can be considered as generalization of an ideal in
a polynomial ring. Just as in the case of an ideal, a Gr\"obner basis
for a module is a generating set with special properties. For the
module under consideration we obtain a Gr\"obner basis using \break
{\it Macaulay 2 }: 
\bea
G^{(1)} &=& (- \D_1, - \D_2, - \D_3, \D_1, \D_2, \D_3) \ , \no \\
G^{(2)} &=& (\D_2-\D_1 \D_3, 0, 1 - \D_3^2, 0, \D_2 \D_3 - \D_1, \D_3^2 - 1) \ , \no \\
G^{(3)} &=& (- \D_1 \D_2, -1, - \D_1, \D_3, 1, \D_2 \D_3) \ , \no \\
G^{(4)} &=& (-1, -\D_3, -\D_1 \D_3, 1, \D_1 \D_2, \D_2) \ , \no \\
G^{(5)} &=& (\D_3 (1 - \D_1^2), \D_3^2 - 1, 0, 0, 1 - \D_1^2, \D_1
(\D_3^2 - 1)) \ .
\label{grbn6}
\eea  
Note that in this Gr\"obner basis $G^{(1)} = \zeta = X^{(2)}, G^{(2)} = 
X^{(1)},  G^{(3)} = \beta = X^{(4)}, G^{(4)} = \alpha = X^{(3)}$. 
Only $G^{(5)}$ is the new generator. 
\par
Another set of generators are just $\alpha, \beta, \gamma$ and $\zeta$. 
This can be checked using {\it Macaulay 2} or one can relate $\alpha, 
\beta, \gamma$ and $\zeta$ to the generators $X^{(A)}, A = 1, 2, 3, 4$ by 
polynomial matrices. In Appendix B, we express the 7 generators we obtained 
following the literature, in terms of  $\alpha, \beta, \gamma$ and $\zeta$. 
Also we express $\alpha, \beta, \gamma$ and $\zeta$  in terms of $X^{(A)}$. 
This proves that all these sets generate the  required module of syzygies.
\par
The question now arises as to which set of generators we should choose
which facilitates further analysis. The analysis is simplified if we
choose a smaller number of generators. Also we would prefer low degree
polynomials to appear in the generators so as to avoid cancellation of
leading terms in the polynomials. By these two criteria we may choose,
$X^{(A)}$ or $\alpha, \beta, \gamma, \zeta$. However, $\alpha, \beta,
\gamma, \zeta$ possess the additional property that this set is left
invariant under a cyclic permutation of indices $1,2,3$. It is found
that this set is more convenient to use because of this symmetry.

\subsection{Canceling optical bench motion noise}

There are now twelve Doppler data streams which have to be combined in
an appropriate manner in order to cancel the noise from the laser as
well as from the motion of the optical benches. As in the previous
case of cancelling laser phase noise, here too, we keep the relevant
terms only, namely, those terms containing laser phase noise and
optical bench motion noise. We then have the following expressions for
the four data streams on spacecraft $1$: 
\bea 
s_1 &=& \D_3 [p_2' + \nu_0 \un_3 \cdot {\vec \Delta}_2' ]
- [ p_1 - \nu_0 \un_3 \cdot {\vec \Delta}_1 ] \ , 
\label{s1}
\\
s_1' &=& \D_2 [p_3 - \nu_0 \un_2 \cdot {\vec \Delta}_3 ]
- [p_1' + \nu_0 \un_2  \cdot  {\vec \Delta}_1' ] \ , 
\label{s1'}
\\
\tau_1 &=& p_1' - p_1 + 2 \nu_0 \un_2 \cdot {\vec \Delta}_1' + \mu_1 \ ,  
\label{tau1}
\\
\tau_1' &=& p_1 - p_1' - 2 \nu_0 \un_3 \cdot {\vec \Delta}_1 + \mu_1
\label{tau1'}
\, .  
\eea 
The other eight data streams on spacecraft 2 and 3 are
obtained by cyclic permutations of the indices in the above equations.
In order to simplify the derivation of the expressions cancelling the optical
bench noises, we note that by subtracting eq. (\ref{tau1'}) from
eq. (\ref{tau1}), we can rewriting the resulting expression (and those
obtained from it by permutation of the spacecraft indices) in the
following form,
\begin{equation}
z_1 \equiv \frac{1}{2} (\tau_1 - \tau_1') = \phi_1' - \phi_1 \, ,
\label{z1}
\end{equation}
where $\phi_1'$, $\phi_1$ are defined as,
\bea
\phi_1' &\equiv& p_1' + \nu_0 \un_2 \cdot {\vec \Delta}_1' \ , 
\no \\
\phi_1 &\equiv& p_1 - \nu_0 \un_3 \cdot {\vec \Delta}_1 \, ,
\eea 
The importance in defining these combinations is that the expressions
for the data streams $s_i$, $s_i'$ simplify into the
following form,
\bea
s_1 &=& \D_3 \phi_2' - \phi_1 \, , \no \\
s_1' &=& \D_2 \phi_3 - \phi_1' \, .
\eea 
If we now combine the $s_i$, $s_i'$, and $z_i$ in the following way,
\begin{eqnarray}
\eta_{1}  & \equiv &  s_{1} - \D_3 z_2
= \D_3 \phi_2 - \phi_1
\ , \ 
\eta_{1'} \equiv s_{1'} + z_1 
= \D_{2} \phi_3 - \phi_1 \ ,
\label{eq:novea}
\\
\eta_{2}  & \equiv &  s_{2} - \D_1 z_3
= \D_1 \phi_3 - \phi_2
\ , \ 
\eta_{2'} \equiv s_{2'} + z_2 
= \D_{3} \phi_1 - \phi_2 \ ,
\label{eq:noveb}
\\
\eta_{3}  & \equiv &  s_{3} - \D_2 z_1
= \D_2 \phi_1 - \phi_3
\ , \ 
\eta_{3'} \equiv s_{3'} + z_3 
= \D_{1} \phi_2 - \phi_3 \ ,
\label{eq:novec}
\end{eqnarray}
we have just reduced the problem of cancelling of six laser and six
optical bench noises to the equivalent problem of removing the three
random processes, $\phi_1$, $\phi_2$, and $\phi_3$, from the six
linear combinations, $\eta_i$, $\eta_i'$, of the one-way measurements
$s_i$, $s_i'$, and $z_i$. By comparing the equations above to equation
(\ref{beams}) for the simpler configuration with only three lasers,
analyzed in the previous section, we see that they are identical in
form.

\subsection{Physical Interpretation of the TDI combinations}

It is important to notice that the four interferometric
combinations ($\alpha, \beta, \gamma, \zeta$), which can be used as a
basis for generating the entire TDI space, are actually synthesized Sagnac
interferometers. This can be seen by rewriting the expression for
$\alpha$, for instance, in the following form,
\begin{equation}
\alpha = [\eta_{1'} + \D_{2} \eta_{3'} + \D_{1} \D_{2'} \eta_{2'}]
-
[\eta_{1} + \D_{3} \eta_{2} + \D_{1} \D_{3} \eta_{2}] 
\, ,
\label{alpha_Phys}
\end{equation}
and noticing that the first square bracket on the right-hand side of
equation (\ref{alpha_Phys}) contains a combination of one-way
measurements describing a light beam propagating clockwise around the
array, while the other terms in the second square-bracket give the
equivalent of another beam propagating counter-clockwise around the
constellation.

Contrary to $\alpha$, $\beta$, and $\gamma$, $\zeta$ can not be
visualized as the difference(or interference) of two synthesized
beams. However, it should still be regarded as a Sagnac combination 
since there exists a  time-delay relationship between it and
$\alpha$, $\beta$, and $\gamma$ \cite{AET99}, 
\begin{equation}
\zeta - \D_1 \D_2 \D_3 \zeta = \D_1 \alpha - \D_2 \D_3 \alpha +
\D_2 \alpha - \D_3 \D_1 \beta + \D_3 \gamma - \D_1 \D_2 \gamma \, .
\label{zeta_all}
\end{equation}
As a consequence of the time-structure of this relationship, $\zeta$
has been called the {\it Symmetrized Sagnac} combination. 

By using the four generators, it is possible to construct several
other interferometric combinations, such as the unequal-arm Michelson
($X, Y, Z$), the Beacons ($P, Q, R$), the Monitors ($E, F, G$), and
the Relays ($U, V, W$).  Contrary to the Sagnac combinations, these
only use four of the six data combinations $\eta_i$,
$\eta_i'$. For this reason they have obvious utility in the event of
selected subsystem failures \cite{ETA00}. 

\begin{figure}
\centering
\includegraphics[width=6in, angle=0]{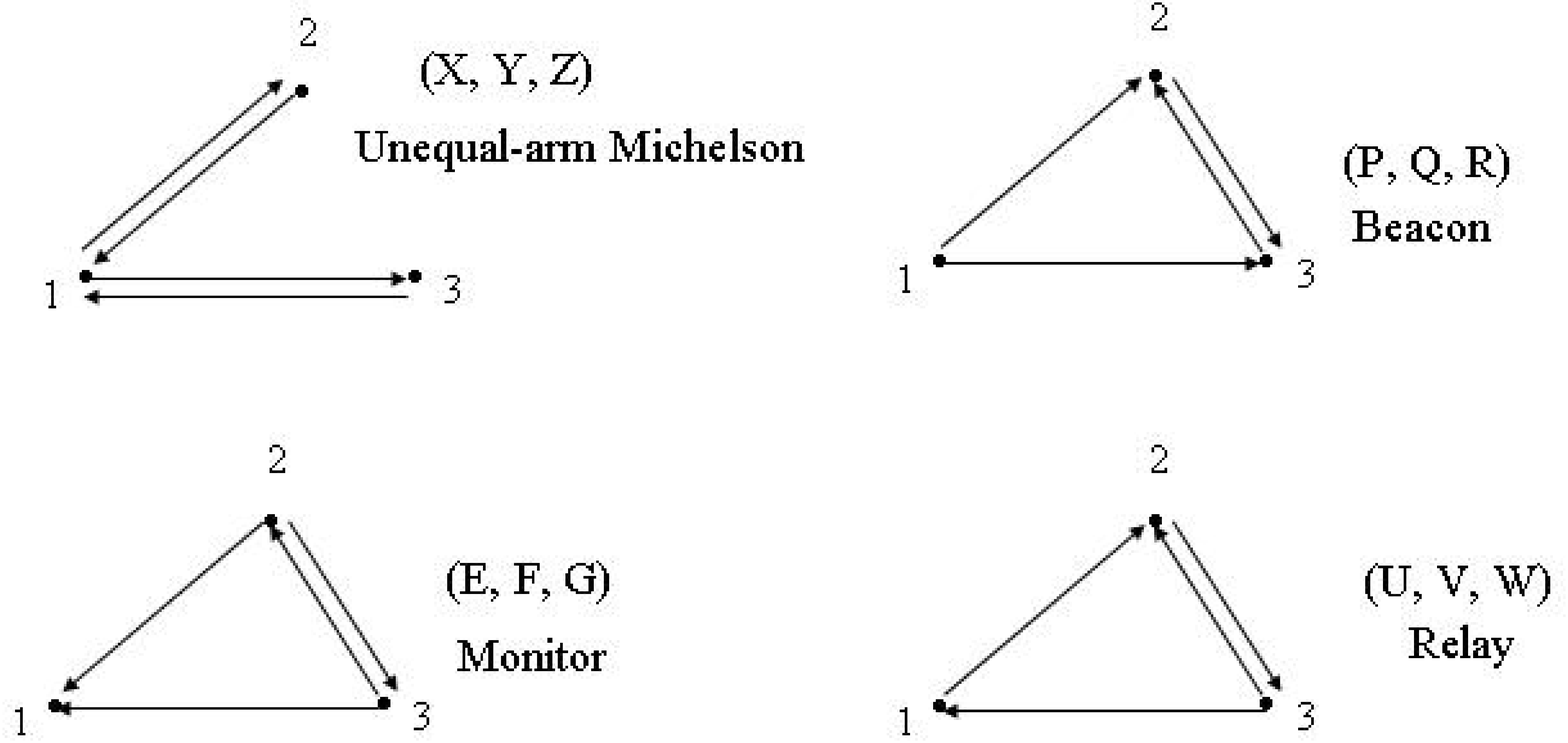}
\caption{Schematic diagrams of the unequal-arm Michelson, Monitor,
  Beacon, and Relay combinations. These TDI combinations rely only on
  four of the six one-way Doppler measurements, as illustrated here.}
\label{All}
\end{figure}

These observables can be written in terms of the Sagnac observables
($\alpha, \beta, \gamma, \zeta$) in the following way, 
\begin{eqnarray}
\D_1 X & = & \D_2 \D_3 \alpha - \D_2 \beta - \ D_3 \gamma + \zeta \ ,
\no \\
P & = & \zeta - \D_1 \alpha \ ,
\no \\
E & = & \D_1 - \D_1 \zeta \ ,
\no \\
U & = & \D_1 \gamma - \beta \ ,
\label{relationships}
\end{eqnarray}
as it is easy to verify by substituting the expressions for the Sagnac
combinations into the above equations. Their physical interpretations
are schematically shown in Figure \ref{All}. 
In the case of the combination $X$, in particular, by writing it
in the following form \cite{AET99},
\begin{equation}
X  = [(\eta_1' + \D_{2'}\eta_3) + \D_{2'} \D_2 (\eta_{1} + \D_3 \eta_2')]
- [(\eta_{1} + \D_3 \eta_2') + \D_3 \D_{3'} (\eta_1' + \D_{2'} \eta_{3})] \ ,
\label{Xcomb}
\end{equation}
one can notice (as pointed out in \cite{Summers} and \cite{STEA03})
that this combination can be visualized as the difference of two sums
of phase measurements, each corresponding to a specific light path
from a laser onboard spacecraft $1$ having phase noise $\phi_1$. The
first square-bracket term in equation (\ref{Xcomb}) represents a
synthesized light-beam transmitted from spacecraft $1$ and made to
bounce once at spacecraft $2$ and $3$ respectively. The second
square-bracket term instead corresponds to another beam also
originating from the same laser, experiencing the same overall delay
as the first beam, but bouncing off spacecraft $3$ first and then
spacecraft $2$. When they are recombined they will cancel the laser
phase fluctuations exactly, having both experienced the same total
delay (assuming stationary spacecraft). The $X$ combinations should
therefore be regarded as the response of a zero-area Sagnac
interferometer.

\section{Time-Delay Interferometry with moving spacecraft}
\label{SECV}

The rotational motion of the LISA array results in a difference of the
light travel times in the two directions around a Sagnac circuit
\cite{S03},\cite{CH03}.  Two time delays along each arm must be used, say
$L_i^{'}$ and $L_i$ for clockwise or counterclockwise propagation as
they enter in any of the TDI combinations. Furthermore, since $L_i$
and $L_i^{'}$ not only differ from one another but can be time
dependent (they "flex"), it was shown that the ``first generation'' TDI
combinations do not completely cancel the laser phase noise (at least
with present laser stability requirements), which can enter at a level
above the secondary noises.  For LISA, and assuming $\dot L_i \simeq
10 {\rm m/sec}$ \cite{Folkner}, the estimated magnitude of the
remaining frequency fluctuations from the laser can be about $30$
times larger than the level set by the secondary noise sources in the
center of the frequency band. In order to solve this potential
problem, it has been shown that there exist new TDI combinations that
are immune to first order shearing (flexing, or constant rate of
change of delay times). These combinations can be derived by using the
time-delay operators formalism introduced in the previous section,
although one has to keep in mind that now these operators no longer
commute \cite{TEA04}. 

In order to derive the new, ``flex-free'' TDI combinations we will
start by taking specific combinations of the one-way data entering in
each of the expressions derived in the previous section. These
combinations are chosen in such a way so as to retain only one of the three
noises $\phi_i , i=1, 2, 3$ if possible.  In this way we can then
implement an iterative procedure based on the use of these basic
combinations and of time-delay operators, to cancel the laser noises
after dropping terms that are quadratic in $\dot{L}/c$ or linear in
the accelerations. This iterative time-delay method, to first order in
the velocity, is illustrated abstractly as follows.  Given a function
of time $\Psi = \Psi(t)$, time delay by $L_i$ is now denoted either
with the standard comma notation \cite{AET99} or by applying the delay
operator $\D_{i}$ introduced in the previous section
\begin{equation}
\D_{i} \Psi = \Psi_{,i} \equiv \Psi(t - L_i(t)) \ . 
\label{eq:25}
\end{equation}
\noindent
We then impose a second time delay $L_j(t)$:
\begin{eqnarray}
\D_{j} \D_{i} \Psi = \Psi_{;ij} & \equiv &  \Psi(t - L_j(t) - L_i(t - L_j(t)))
\nonumber \\
&  \simeq  & \Psi(t - L_j(t) - L_i(t) + \dot  L_i(t)  L_j) 
\nonumber \\
&  \simeq & \Psi_{,ij} + \dot \Psi_{,ij} \dot L_i L_j \ .
\label{eq:26}
\end{eqnarray}

\noindent
A third time delay $L_k(t)$ gives:
\begin{eqnarray}
\D_{k} \D_{j} \D_{i} \Psi = \Psi_{;ijk} & = & 
\Psi(t - L_k(t) - L_j(t - L_k(t)) - L_i(t - L_k(t) - L_j(t - L_k(t))))
\nonumber \\
& & \simeq \Psi_{,ijk} + \dot \Psi_{,ijk} [\dot L_i (L_j + L_k) + \dot L_j L_k] \ ,
\label{eq:27}
\end{eqnarray}
\noindent
and so on, recursively; each delay generates a first-order correction
proportional to its rate of change times the sum of all delays coming
after it in the subscripts. Commas have now been replaced with
semicolons \cite{STEA03}, to remind us that we consider moving arrays.
When the sum of these corrections to the terms of a data combination
vanishes, the combination is called flex-free.

Also, note that each delay operator, $\D_{i}$, has a unique inverse,
$D^{-1}_{i}$, whose expression can be derived by requiring that
$D^{-1}_{i} \D_{i} = I$, and neglecting quadratic and higher order
velocity terms. Its action on a time series $\Psi (t)$ is
\begin{equation}
D^{-1}_{i} \Psi (t) \equiv \Psi (t + L_i (t + L_i)) \ .
\label{eq:28}
\end{equation}
\noindent
Note that this is not like an advance operator one might expect,
since it advances not by $L_i (t)$ but rather $L_i (t + L_i)$.

\subsection{The Unequal-Arm Michelson}

The unequal-arm Michelson combination relies on the four measurements
$\eta_{1}$, $\eta_{1'}$, $\eta_{2'}$, and $\eta_{3}$.  Note that the
two combinations $\eta_{1} + \eta_{2',3}$, $\eta_{1'} + \eta_{3,2'}$
represent the two synthesized two-way data measured onboard
spacecraft $1$, and can be written in the following form
\begin{eqnarray}
\eta_{1} + \eta_{2',3} & = & \left(\D_{3}\D_{3'} - I\right) \ \phi_1 \ ,
\label{eq:diecia}
\\
\eta_{1'} + \eta_{3,2'}  & = & \left(\D_{2'}\D_{2} - I\right) \ \phi_1 \ ,
\label{eq:diecib}
\end{eqnarray}
where $I$ is the identity operator. Since in the
stationary case any pairs of these operators commute, i.e.  $\D_i
\D_{j'} - \D_{j'} \D_i = 0$, from equations (\ref{eq:diecia},
\ref{eq:diecib}) it is easy to derive the following expression for the
unequal-arm interferometric combination, $X$, which eliminates, 
$\phi_1$
\begin{equation}
X =  \left[\D_{2'}\D_{2} - I\right] (\eta_{1} + \eta_{2',3}) - 
\left[\left(\D_{3}\D_{3'} - I\right)\right] (\eta_{1'} + \eta_{3,2'}).
\label{eq:undici}
\end{equation}
\noindent
If, on the other hand, the time-delays depend on time, the expression
of the unequal-arm Michelson combination above no longer cancels
$\phi_1$. In order to derive the new expression for the unequal-arm
interferometer that accounts for ``flexing'', let us first consider
the following two combinations of the one-way measurements entering
into the $X$ observable given in equation (\ref{eq:undici}):
\begin{eqnarray}
\left[(\eta_{1'} + \eta_{3;2'}) + (\eta_{1} + \eta_{2;3})_{;22'}\right] & = &  
\left[D_{2'}D_{2}D_{3}D_{3'} - I \right] \phi_1 \ ,
\label{eq:ventinovea}
\\
\left[(\eta_{1} + \eta_{2';3}) + (\eta_{1'} + \eta_{3;2'})_{;3'3}\right] & = &
\left[D_{3}D_{3'}D_{2'}D_{2} - I \right] \phi_1 \ .
\label{eq:ventinoveb}
\end{eqnarray}
Using equations (\ref{eq:ventinovea}, \ref{eq:ventinoveb}) we can use the delay
technique again to finally derive the following expression for the new
unequal-arm Michelson combination $X_1$ that accounts for the flexing
effect,
\begin{eqnarray}
X_1 & = & \left[D_{2}D_{2'}D_{3'}D_{3} - I \right] \ 
\left[(\eta_{21} + \eta_{12;3'}) + (\eta_{31} +
  \eta_{13;2})_{;33'}\right]
\nonumber
\\
& & -
\left[D_{3'}D_{3}D_{2}D_{2'} - I \right] \ 
\left[(\eta_{31} + \eta_{13;2}) + (\eta_{21} + \eta_{12;3'})_{;2'2}\right].
\label{eq:trenta}
\end{eqnarray}

\noindent
As usual, $X_2$ and $X_3$ are obtained by cyclic permutation of the
spacecraft indices. This expression is readily shown to be laser-noise-free
to first order of spacecraft separation velocities $\dot L_i$: it is
``flex-free''.

\subsection{The Sagnac Combinations}

In the above subsection we have used the same symbol $X$ for the
unequal-arm Michelson combination for both the rotating (i.e.
constant delay times) and stationary cases.  This emphasizes that, for
this TDI combination (and, as we will see below, also for all the
combinations including only four links) the forms of the equations do
not change going from systems at rest to the rotating case.  One needs
only distinguish between the time-of-flight variations in the
clockwise and counter-clockwise senses (primed and unprimed delays).

In the case of the Sagnac variables, ($\alpha, \beta, \gamma, \zeta$),
however, this is not the case as it is easy to understand on simple
physical grounds. In the case of $\alpha$ for instance, light
originating from spacecraft 1 is simultaneously sent around the array
on clockwise and counterclockwise loops, and the two returning beams
are then recombined. If the array is rotating, the two beams
experience a different delay (the Sagnac effect), preventing the noise
$\phi_1$ from cancelling in the $\alpha$ combination. 

In order to find the solution to this problem let us first rewrite
$\alpha$ in such a way to explicitly emphasize what it does:
attempts to remove the same fluctuations affecting two beams that have
been made to propagated clockwise and counter-clockwise around the
array,
\begin{equation}
\alpha = [\eta_{1'} + \D_{2'} \eta_{3'} + \D_{1'} \D_{2'} \eta_{2'}]
-
[\eta_{1} + \D_{3} \eta_{2} + \D_{1} \D_{3} \eta_{2}]
\, ,
\label{alpha}
\end{equation}
where we have accounted for clockwise and counterclockwise light
delays. It is straightforward to verify that this combination no
longer cancels the laser and optical bench noises.  If, however, we
expand the two terms inside the square-brackets on the right-hand-side
of equation (\ref{alpha}) we find that they are equal to:
\begin{equation}
[\eta_{1'} + \D_{2'} \eta_{3'} + \D_{1'} \D_{2'} \eta_{2'}] = 
[\D_{2'}\D_{1'}\D_{3'} - I] \phi_1
\nonumber
\end{equation}
\begin{equation}
[\eta_{1} + \D_{3} \eta_{2} + \D_{1} \D_{3} \eta_{2}] =
[\D_{3}\D_{1}\D_{2} - I] \phi_1 \, .
\label{alpha_int}
\end{equation}
If we now apply our iterative scheme to the combinations given in
equation (\ref{alpha_int}) we finally get the expression for the
Sagnac combination, $\alpha_1$, that is unaffected by laser noise in
presence of rotation,
\begin{equation}
\alpha_1  = [\D_{3}\D_{1}\D_{2} - I] \ 
[\eta_{1'} + \D_{2'} \eta_{3'} + \D_{1'} \D_{2'} \eta_{2'}] 
- [\D_{2'}\D_{1'}\D_{3'} - I] 
[\eta_{1} + \D_{3} \eta_{2} + \D_{1} \D_{3} \eta_{2}] \ .
\label{alpha1}
\end{equation}
If the delay-times are also time-dependent, we find that the residual
laser noise remaining into the combination $\alpha_1$ is actually
equal to
\begin{eqnarray}
\dot \phi_{1,1231'2'3'} 
[(\dot L_1 + \dot L_2 + \dot L_3) (L_1^{'} + L_2^{'} + L_3^{'})
- (\dot L_1^{'} + \dot L_2^{'} + \dot L_3^{'}) (L_1 + L_2 + L_3)] \ .
\end{eqnarray}
Fortunately, although first order in the relative velocities, the
residual is small, as it involves the difference of the clockwise and
counterclockwise rates of change of the propagation delays on the
{\it{same}} circuit.  For LISA, the remaining laser phase noises in
$\alpha_i$, $i = 1, 2, 3$, are several orders of magnitude below the
secondary noises.

In the case of $\zeta$, however, the rotation of the array breaks the
symmetry and therefore its uniqueness. However, there still exist
three generalized TDI laser-noise-free data combinations that have
properties very similar to $\zeta$, and which can be used for the same
scientific purposes \cite{TAE01}.  These combinations, which we call
($\zeta_1, \zeta_2, \zeta_3$), can be derived by applying again our
time-delay operator approach. 

\noindent
Let us consider the following combination of the $\eta_{i}$, $\eta_{i'}$
measurements, each being delayed only once \cite{AET99}:
\begin{eqnarray}
\eta_{3,3} - \eta_{3',3} + \eta_{1,1'} & = & \left[D_{3} D_{2} -
  D_{1'}\right] \phi_1 \ ,
\label{eq:ventiduea}
\\
\eta_{1',1} - \eta_{2,2'} + \eta_{2',2'} & = & \left[D_{3'} D_{2'} -
  D_{1}\right] \phi_1 \ ,
\label{eq:ventidueb}
\end{eqnarray}
where we have used the commutativity property of the delay operators
in order to cancel the $\phi_2$ and $\phi_3$ terms. Since both sides
of the two equations above contain only the $\phi_1$ noise, 
$\zeta_1$ is found by the following expression:
\begin{equation}
\zeta_1 = \left[D_{3'}D_{2'} - D_{1}\right] \left(\eta_{31,1'} - \eta_{32,2} + \eta_{12,2}\right)
- \left[D_{2}D_{3} - D_{1'}\right]\left(\eta_{13,3'} - \eta_{23,3'} +
  \eta_{21,1}\right) \ .
\label{eq:ventitre}
\end{equation}
If the light-times in the arms are equal in the clockwise
and counterclockwise senses (e.g. no rotation) there is no distinction
between primed and unprimed delay times.  In this case, $\zeta_1$ is
related to our original symmetric Sagnac $\zeta$ by $\zeta_1 =
\zeta_{,23} - \zeta_{,1}$.  Thus for the practical LISA case (arm
length difference $< 1 \%$), the SNR of $\zeta_1$ will be the same as
the SNR of $\zeta$.

If the delay-times also change with time, the perfect cancellation of
the laser noises is no longer achieved in the ($\zeta_1, \zeta_2,
\zeta_3$) combinations. However, it has been shown in \cite{TEA04}
that the magnitude of the residual laser noises in these combinations
are significantly smaller than the LISA secondary system noises,
making their effects entirely negligible.

The expressions for the Monitor, Beacon, and Relay combinations,
accounting for the rotation and flexing of the LISA array, have been
derived in the literature \cite{TEA04} by applying the time-delay
iterative procedure highlighted in this section. The interested reader is
referred to that paper for details.

A mathematical formulation of the "second generation" TDI, which
generalizes the one presented in Section \ref{SECIV} for the
stationary LISA, still needs to be derived.  In the case when only the
Sagnac effect is considered (and the delay-times remain constant in
time) the mathematical formulation of Section \ref{SECIV} can be
extended in a straight forward way where now the six time-delays
$\D_{i}$ and $\D_{i}'$ must be taken into account. The polynomial ring
is now in these six variables and the corresponding module of syzygies
can be constructed over this enlarged polynomial ring \cite{NV04}.
However, when the arms are allowed to flex, that is, the operators
themselves are functions of time, the operators no longer commute. One
must then resort to non-commutative Gr\"obner basis methods. We will
investigate this mathematical problem in the near future.

\section{Optimal LISA Sensitivity}
\label{SECVI}

All the above interferometric combinations have been shown to
individually have rather different sensitivities \cite{ETA00}, as a
consequence of their different responses to gravitational radiation
and system noises. Since LISA has the capability of {\underbar
  {simultaneously}} observing a gravitational wave signal with many
different interferometric combinations (all having different antenna
patterns and noises),  we should no longer regard LISA as a single
detector system but rather as an array of gravitational wave detectors
working in coincidence. This suggests that the presently adopted LISA
sensitivity could be improved by {\it optimally} combining elements of
the TDI space. 

Before proceeding with this idea, however, let us consider again the
so called ``second generation'' TDI Sagnac observables: ($\alpha_1,
\alpha_2, \alpha_3$). The expressions of the gravitational wave signal
and the secondary noise sources entering into $\alpha_1$ will in
general be different from those entering into $\alpha$, the
corresponding Sagnac observable derived under the assumption of a
stationary LISA array \cite{AET99}, \cite{ETA00}. However, the other
remaining, secondary noises in LISA are so much smaller, and the
rotation and systematic velocities in LISA are so intrinsically small,
that index permutation may still be done for them \cite{TEA04}. It is
therefore easy to derive the following relationship between the signal
and secondary noises in $\alpha_1$, and those entering into the
stationary TDI combination $\alpha$ \cite{STEA03}, \cite{TEA04}:
\begin{equation}
\alpha_1 (t) \simeq \alpha (t) - \alpha (t - L_1 - L_2 - L_3) \ ,
\label{eq:duebis}
\end{equation}
where $L_i  \ , \ \ i=1, 2, 3$ are the unequal-arm lengths of the
stationary LISA array. Equation (\ref{eq:duebis}) implies that any data
analysis procedure and algorithm that will be implemented for the
second-generation TDI combinations can actually be derived by
considering the corresponding ``first generation'' TDI combinations.
For this reason, from now on we will focus our attention on the
gravitational wave responses of the first-generation TDI observables
($\alpha, \beta, \gamma, \zeta$).

As a consequence of these considerations, we can still regard ($\alpha,
\beta, \gamma, \zeta$) as the generators of the TDI space, and write
the most general expression for an element of the TDI space, $\eta
(f)$, as a linear combination of the Fourier transforms of the four
generators (${\widetilde {\alpha}}, {\widetilde {\beta}}, {\widetilde
  {\gamma}}, {\widetilde {\zeta}}$)
\begin{equation}
\eta(f) \equiv a_1 (f, {\vec \lambda}) \ {\widetilde{\alpha}} (f) \ +
\ a_2 (f, {\vec \lambda}) \ {\widetilde{\beta}} (f) \ +
\ a_3 (f, {\vec \lambda}) \ {\widetilde{\gamma}} (f) \ +
\ a_4 (f, {\vec \lambda}) \ {\widetilde{\zeta}} (f)  \ ,
\label{eq:otto}
\end{equation}
where the $\{a_i (f, \vec \lambda)\}^4_{i=1}$ are arbitrary complex
functions of the Fourier frequency $f$, and of a vector $\vec \lambda$
containing parameters characterizing the gravitational wave signal
(source location in the sky, waveform parameters, etc.) and the noises
affecting the four responses (noise levels, their correlations, etc.).
For a given choice of the four functions $\{a_i \}^4_{i=1}$, $\eta$
gives an element of the functional space of interferometric
combinations generated by ($\alpha, \beta, \gamma, \zeta$). Our goal
is therefore to identify, for a given gravitational wave signal, the
four functions $\{a_i \}^4_{i=1}$ that maximize the signal-to-noise
ratio, ${\rm SNR}_{\eta}^2$, of the combination $\eta$,
\begin{equation}
{\rm SNR}_{\eta}^2 = 
\int_{f_{l}}^{f_u} 
\frac
{|a_1 \ {\widetilde \alpha_s} + a_2 \ {\widetilde \beta_s} + a_3 \
  {\widetilde \gamma_s} + a_4 {\widetilde \zeta_s} |^2}
{\langle|a_1 \ {\widetilde \alpha_n} + a_2 \ {\widetilde \beta_n} +
  a_3 \ {\widetilde \gamma_n} + a_4 {\widetilde \zeta_n} |^2 \rangle}  \ df \ .
\label{eq:9bis}
\end{equation}
In equation (\ref{eq:9bis}) the subscripts $s$ and $n$ refer to the
signal and the noise parts of (${\widetilde {\alpha}}, {\widetilde
  {\beta}}, {\widetilde {\gamma}}, {\widetilde {\zeta}}$)
respectively, the angle brackets represent noise ensemble averages,
and the interval of integration ($f_l, f_u$) corresponds to the
frequency band accessible by LISA.

Before proceeding with the maximization of the ${\rm SNR}_{\eta}^2$ we may
notice from equation (\ref{zeta_all}) that the Fourier transform of the
totally symmetric Sagnac combination, $\widetilde \zeta$, multiplied
by the transfer function $1 - e^{2 \pi i f (L_1 + L_2 + L_3)}$ can be
written as a linear combination of the Fourier transforms of the
remaining three generators ($ {\widetilde {\alpha}}, {\widetilde
  {\beta}}, {\widetilde {\gamma}}$). Since the signal-to-noise ratio
of $\eta$ and $(1 - e^{2 \pi i f (L_1 + L_2 + L_3)}) \eta$ are equal,
we may conclude that the optimization of the signal-to-noise ratio of
$\eta$ can be performed only on the three observables $\alpha, \beta,
\gamma$. This implies the following redefined expression for
${\rm SNR}_{\eta}^2$:
\begin{equation}
{\rm SNR}_{\eta}^2 = 
\int_{f_{l}}^{f_u} 
\frac
{|a_1 \ {\widetilde \alpha_s} + a_2 \ {\widetilde \beta_s} + a_3 \
  {\widetilde \gamma_s} |^2} 
{\langle|a_1 \ {\widetilde \alpha_n} + a_2 \ {\widetilde \beta_n} +
  a_3 \ {\widetilde \gamma_n} |^2 \rangle} \ df \ .
\label{eq:9}
\end{equation}
The ${\rm SNR}_{\eta}^2$ can be regarded as a functional over the space of
the three complex functions $\{ a_i \}^3_{i=1}$, and the particular
set of complex functions that extremize it can of course be derived by
solving the associated set of Euler-Lagrange equations.

In order to make the derivation of the optimal SNR easier, let us
first denote by ${\bf x}^{(s)}$ and ${\bf x}^{(n)}$ the two vectors of
the signals (${\widetilde{\alpha}}_s, {\widetilde{\beta}}_s,
{\widetilde{\gamma}}_s$) and the noises (${\widetilde{\alpha}}_n,
{\widetilde{\beta}}_n, {\widetilde{\gamma}}_n$) respectively. Let us also
define $\bf a$ to be the vector of the three functions $\{a_i
\}^3_{i=1}$, and denote with ${\bf C}$ the hermitian, non-singular,
correlation matrix of the vector random process ${\bf x}_n$,
\begin{equation}
({\bf C})_{rt} \equiv \langle {\bf x}^{(n)}_{r} {\bf x}^{(n)*}_{t} \rangle \ .
\label{eq:14}
\end{equation}
If we finally define $({\bf A})_{ij}$ to be the components of the
hermitian matrix ${\bf x}^{(s)}_i {\bf x}^{(s)*}_j$, we can rewrite
${\rm SNR}_{\eta}^2$ in the following form,
\begin{equation}
{\rm SNR}_{\eta}^2 = 
\int_{f_{l}}^{f_u} 
\frac{{\bf a}_i {\bf A}_{ij} {\bf a}^*_j }
{{\bf a}_r {\bf C}_{rt} {\bf a}^*_t }
 \ df \ ,
\label{eq:16}
\end{equation}
where we have adopted the usual convention of summation over repeated
indices. Since the noise correlation matrix ${\bf C}$ is non-singular,
and the integrand is positive definite or null, the stationary values
of the signal-to-noise ratio will be attained at the stationary values
of the integrand, which are given by solving the following set of
equations (and their complex conjugated expressions):
\begin{equation}
\frac{\partial}{\partial {\bf a}_k} \ \left[
\frac
{{\bf a}_i {\bf A}_{ij} {\bf a}^*_j }
{{\bf a}_r {\bf C}_{rt} {\bf a}^*_t }
 \right] = 0 \ \ \ , \ \ \ k = 1, 2, 3 \ .
\label{eq:16bis}
\end{equation}
After taking the partial derivatives, equation (\ref{eq:16bis}) can be
rewritten in the following form,
\begin{equation}
({\bf C}^{-1})_{ir} 
({\bf A})_{rj} ({\bf a}^*)_j = \left[
\frac
{{\bf a}_p {\bf A}_{pq} {\bf a}^*_q }
{{\bf a}_l {\bf C}_{lm} {\bf a}^*_m }
 \right]  
\ ({\bf a}^*)_i \ \ \ , \ \ \ i = 1, 2, 3 
\label{eq:16tris}
\end{equation}
which tells us that the stationary values of the signal-to-noise ratio
of $\eta$ are equal to the eigenvalues of the the matrix ${\bf C^{-1}
  \cdot A}$. The result in equation (\ref{eq:16bis}) is well known in
the theory of quadratic forms, and it is called the Rayleigh's
principle \cite{Noble69}, \cite{Selby64}.

In order now to identify the eigenvalues of the matrix ${\bf C^{-1}
  \cdot A}$, we first notice that the $3 \times 3$ matrix ${\bf A}$
has rank $1$. This implies that the matrix ${\bf
  C}^{-1} \cdot {\bf A}$ has also rank $1$, as it is easy to verify.
Therefore two of its three eigenvalues are equal to zero, while the
remaining non-zero eigenvalue represents the solution we are looking
for.

The analytic expression of the third eigenvalue can be obtained by
using the property that the trace of the $3 \times 3$ matrix ${\bf
  C}^{-1} \cdot {\bf A}$ is equal to the sum of its three eigenvalues,
and in our case to the eigenvalue we are looking for.  From these
considerations we derive the following expression for the optimized
signal-to-noise ratio ${{\rm SNR}_{\eta}^2}_{\rm opt.}$:
\begin{equation}
{{\rm SNR}_{\eta}^2}_{\rm opt.} = \int_{f_{l}}^{f_u} 
{\bf x}^{(s)*}_i \ ({\bf C}^{-1})_{ij} \ {\bf x}^{(s)}_j  \ df \ .
\label{eq:18}
\end{equation}
\noindent
We can summarize the results derived in this section, which are given
by equations (\ref{eq:9},\ref{eq:18}), in the following way:

\noindent
(i) among all possible interferometric combinations LISA will be able
to synthesize with its four generators $\alpha, \beta, \gamma, \zeta$,
the particular combination giving maximum signal-to-noise ratio can be
obtained by using only three of them, namely ($\alpha, \beta,
\gamma$);

\noindent
(ii) the expression of the optimal signal-to-noise ratio given by
equation (\ref{eq:18}) implies that LISA should be regarded as a
network of three interferometer detectors of gravitational radiation
(of responses ($\alpha, \beta, \gamma$)) working in coincidence
\cite{Finn01, NPDV03_1}.

\subsection{General application}
\label{GA}

As an application of equation (\ref{eq:18}), here we calculate the
sensitivity that LISA can reach when observing sinusoidal signals
uniformly distributed on the celestial sphere and of random
polarization. In order to calculate the optimal signal-to-noise ratio
we will also need to use a specific expression for the noise
correlation matrix ${\bf C}$. As a simplification, we will assume the
LISA arm-lengths to be equal to its nominal value $L = 16.67 \ {\rm
  sec.}$, the optical-path noises to be equal and uncorrelated to each
other, and finally the noises due to the proof-mass noises to be also
equal, uncorrelated to each other and to the optical-path noises.
Under these assumptions the correlation matrix becomes real, its three
diagonal elements are equal, and all the off-diagonal terms are equal
to each other, as it is easy to verify by direct calculation
\cite{ETA00}. The noise correlation matrix ${\bf C}$ is therefore
uniquely identified by two real functions, $S_{\alpha}$ and $S_{\alpha
  \beta}$, in the following way
\[ {\bf C} = \left( \begin{array}{ccc}
S_{\alpha} & S_{\alpha \beta} & S_{\alpha \beta} \\
S_{\alpha \beta} & S_{\alpha} & S_{\alpha \beta} \\
S_{\alpha \beta} & S_{\alpha \beta} & S_{\alpha} 
\end{array} 
\label{eq:19}
\right).\] 

The expression of the optimal signal-to-noise ratio assumes a rather
simple form if we diagonalize this correlation matrix by properly
``choosing a new basis''. There exists an orthogonal
transformation of the generators ($ {\widetilde {\alpha}}, {\widetilde
  {\beta}}, {\widetilde {\gamma}}$) which will transform the optimal
signal-to-noise ratio into the sum of the signal-to-noise ratios of
the ``transformed'' three interferometric combinations.  The expressions
of the three eigenvalues $\{\mu_i\}^3_{i=1}$ (which are real) of the
noise correlation matrix ${\bf C}$ can easily be found by using the
algebraic manipulator {\it Mathematica} \cite{Wolf02}, and they are
equal to
\begin{equation}
\mu_1 = \mu_2 = S_{\alpha} - S_{\alpha \beta} \ \ \ , \ \ \ 
\mu_3 = S_{\alpha} + 2 \ S_{\alpha \beta} \ .
\label{eq:20}
\end{equation}
Note that two of the three real eigenvalues, ($\mu_1, \mu_2$), are
equal. This implies that the eigenvector associated to $\mu_3$ is
orthogonal to the two-dimensional space generated by the eigenvalue
$\mu_1$, while any chosen pair of eigenvectors corresponding to
$\mu_1$ will not necessarily be orthogonal.  This inconvenience can
be avoided by choosing an arbitrary set of vectors in this
two-dimensional space, and by ortho-normalizing them. After some simple
algebra, we have derived the following three ortho-normalized
eigenvectors
\begin{equation}
{\bf v_1} = \frac{1}{\sqrt{2}} \ (-1, 0, 1) \ \ \ , \ \ \ 
{\bf v_2} = \frac{1}{\sqrt{6}} \ (1, -2, 1) \ \ \ , \ \ \ 
{\bf v_3} = \frac{1}{\sqrt{3}} \ (1, 1, 1) \ .
\label{eq:21}
\end{equation}
Equation (\ref{eq:21}) implies the following three linear combinations of the
generators (${\widetilde{\alpha}}, {\widetilde{\beta}},
{\widetilde{\gamma}}$)
\begin{equation}
A = \frac{1}{\sqrt{2}} \ ({\widetilde{\gamma}} - {\widetilde{\alpha}}) \ \ \ , \ \ \ 
E = \frac{1}{\sqrt{6}} \ ({\widetilde{\alpha}} - 2 {\widetilde{\beta}} + {\widetilde{\gamma}}) \ \ \ , \ \ \ 
T = \frac{1}{\sqrt{3}} \ ({\widetilde{\alpha}} + {\widetilde{\beta}} +
{\widetilde{\gamma}}) \ ,
\label{eq:22}
\end{equation}
where $A$, $E$, and $T$ are italicized to indicate that these are
``orthogonal modes''. Although the expressions for the modes $A$ and
$E$ depend on our particular choice for the two eigenvectors (${\bf
  v_1}, {\bf v_2}$), it is clear from our earlier considerations that
the value of the optimal signal-to-noise ratio is unaffected by such a
choice. From equation (\ref{eq:22}) it is also easy to verify that the
noise correlation matrix of these three combinations is diagonal, and
that its non-zero elements are indeed equal to the eigenvalues given
in equation (\ref{eq:20}).

In order to calculate the sensitivity corresponding to the expression
of the optimal signal-to-noise ratio, we have proceeded similarly to
what was done in \cite{AET99}, \cite{ETA00}, and described in more
detail in \cite{WP02}. We assume an equal-arm LISA ($L = 16.67$ light
seconds), and take the one-sided spectra of proof mass and aggregate
optical-path-noises (on a single link), expressed as fractional
frequency fluctuation spectra, to be (\cite{ETA00}, \cite{LISA98}),
$S^{proof\ mass}_y = 2.5 \times {10^{-48}} \ {\left[f/{1 Hz}
  \right]}^{-2} \ {\rm Hz}^{-1}$ and $S^{optical \ path}_y = 1.8
\times {10^{-37}} \ {\left[f/1 Hz \right]}^2 \ {\rm Hz}^{-1}$,
respectively.  We also assume that aggregate optical path noise has
the same transfer function as shot noise.

The optimum SNR is the square root of the sum of the squares of the
SNRs of the three ``orthogonal modes'' ($A, E, T$).  To compare with previous
sensitivity curves of a single LISA Michelson interferometer, we
construct the SNRs as a function of Fourier frequency for sinusoidal
waves from sources uniformly distributed on the celestial sphere. To
produce the SNR of each of the ($A, E, T$) modes we need the
gravitational wave response and the noise response as a function of
Fourier frequency.  We build up the gravitational wave responses of
the three modes ($A, E, T$) from the gravitational wave responses of
($\alpha, \beta, \gamma$).  For $7000$ Fourier frequencies in the
${\sim}10^{-4}$ Hz to ${\sim}1$ Hz LISA band, we produce the Fourier
transforms of the gravitational wave response of ($\alpha, \beta,
\gamma$) from the formulas in \cite{AET99}, \cite{WP02}.  The averaging
over source directions (uniformly distributed on the celestial sphere)
and polarization states (uniformly distributed on the Poincar\'e
sphere) is performed via a Monte Carlo method. From the Fourier
transforms of the ($\alpha, \beta, \gamma$) responses at each
frequency, we construct the Fourier transforms of ($A, E, T$).  We
then square and average to compute the mean-squared responses of
($A, E, T$) at that frequency from $10^4$ realizations of (source
position, polarization state) pairs.

We adopt the following terminology: we refer to a single element of the module
as a data {\it combination}; while a function of the elements of the 
module, such as taking the maximum over several data combinations in the
module or squaring and adding data combinations belonging to the module,  
is called  as an  {\it observable}.  The important point to note is
that the laser frequency noise is also suppressed for the observable although
it may not be an element of the module. 

The noise spectra of ($A, E, T$) are determined from the raw spectra of 
proof-mass and optical-path noises, and the transfer functions of these
noises to ($A, E, T$).  Using the transfer functions given in \cite{ETA00},
the resulting spectra are equal to,
\begin{eqnarray}
S_{A}(f) = S_{E}(f) & = & \ 16 \ \sin^2(\pi f L) \
[3 + 2 \cos(2 \pi f L) + \cos(4 \pi f L)] S^{proof\ mass}_y(f)
\nonumber \\
& + & \ 8 \ \sin^2(\pi f L) \ [2 + \cos(2 \pi f L)] \ S^{optical \
  path}_y(f) \ ,
\end{eqnarray}

\begin{eqnarray}
S_{T}(f) & = & \ 2 [1 + 2 \cos (2 \pi f L)]^2 \ 
[4 \ \sin^2 (\pi f L) S_y^{proof\ mass} + S_y^{optical\ path} (f)] \ .
\end{eqnarray}
Let the amplitude of the sinusoidal gravitational wave be $h$.  The
SNR for, e.g. $A$, ${\rm SNR}_{A}$, at each frequency $f$ is equal to $h$
times the ratio of the root-mean-squared gravitational wave response
at that frequency divided by $\sqrt{S_{A}(f) \ B}$, where $B$ is the
bandwidth conventionally taken to be equal to $1$ cycle per year.
Finally, if we take the reciprocal of ${\rm SNR}_{A}/h$ and multiply it by
$5$ to get the conventional ${\rm SNR} = 5$ sensitivity criterion, we obtain
the sensitivity curve for this combination which can then be compared
against the corresponding sensitivity curve for the equal-arm
Michelson Interferometer.

\begin{figure}
\centering
\includegraphics[width=2.5in, angle=-90]{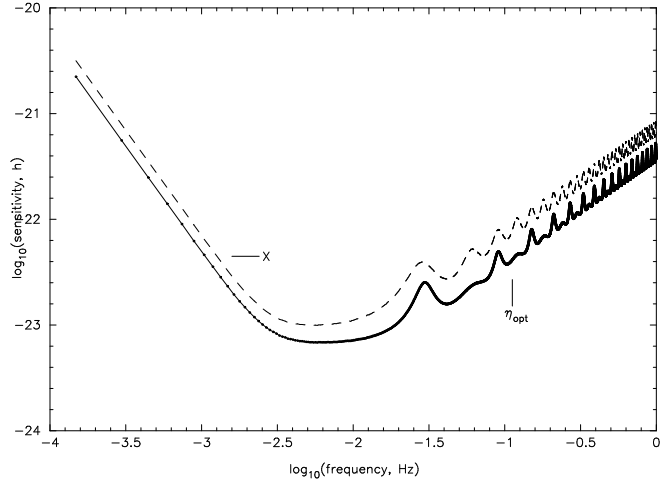}
\caption{The LISA Michelson sensitivity curve (SNR = 5)
and the sensitivity curve for the optimal combination of the data, both
as a function of Fourier frequency. The integration time is equal to
one year, and LISA is assumed to have a nominal armlength $L =
16.67 {\rm sec}$.}
\label{SENS}
\end{figure}

In Figure \ref{SENS} we show the sensitivity curve for the LISA equal-arm
Michelson response (${\rm SNR} = 5$) as a function of the Fourier frequency,
and the sensitivity curve from the optimum weighting of the data
described above: $5 h / \sqrt{{\rm SNR}_{A}^2 + {\rm SNR}_{E}^2 + {\rm SNR}_{T}^2}$. The
SNRs were computed for a bandwidth of 1 cycle/year.  Note that at
frequencies where the LISA Michelson combination has best sensitivity,
the improvement in signal-to-noise ratio provided by the optimal
observable is slightly larger than $\sqrt{2}$.

\begin{figure}
\centering
\includegraphics[width=2.5in, angle=-90]{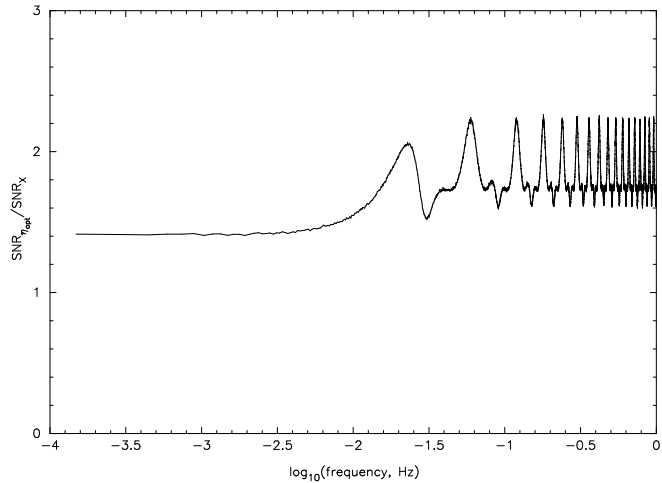}
\caption{The optimal SNR divided by the SNR of a single Michelson 
  interferometer, as a function of the Fourier frequency $f$. The
  sensitivity gain in the low-frequency band is equal to $\sqrt{2}$,
  while it can get larger than $2$ at selected frequencies in the
  high-frequency region of the accessible band. The integration time
  has been assumed to be one year, and the proof mass and optical path
  noise spectra are the nominal ones. See the main body of the paper
  for a quantitative discussion of this point.}
\label{SNRO}
\end{figure}

In Figure \ref{SNRO} we plot the ratio between the optimal SNR and the SNR of a
single Michelson interferometer.  In the long-wavelength limit, the
SNR improvement is $\sqrt{2}$.  For Fourier frequencies greater than
or about equal to $1/L$, the SNR improvement is larger and varies
with the frequency, showing an average value of about $\sqrt{3}$.  In
particular, for bands of frequencies centered on integer multiples of
$1/L$, $SNR_{T}$ contributes strongly and the aggregate SNR in these
bands can be greater than $2$.
\begin{figure}[h]
\centering
\includegraphics[width=2.5in, angle=-90]{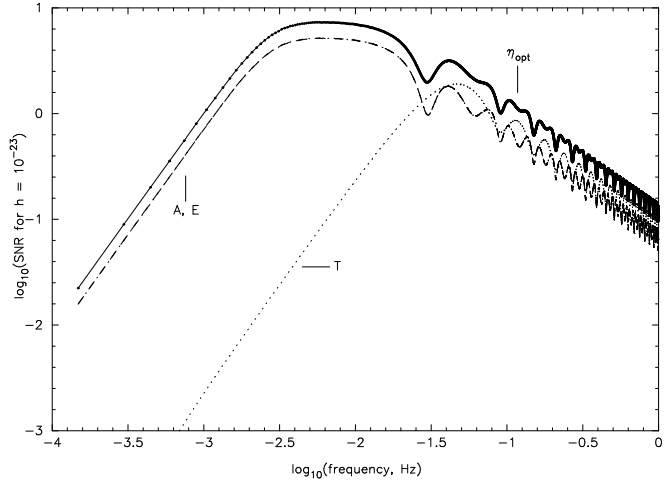}
\caption{The SNRs of the three combinations, ($A, E, T$), and their sum
  as a function of the Fourier frequency $f$. The SNRs of $A$ and $E$
  are equal over the entire frequency band. The SNR of $T$ is
  significantly smaller than the other two in the low part of the
  frequency band, while is comparable to (and at times larger than)
  the SNR of the other two in the high-frequency region. See text for
  a complete discussion.}
\label{OSNR}
\end{figure}

In order to better understand the contribution from the three
different combinations to the optimal combination of the three
generators, in Figure \ref{OSNR} we plot the signal-to-noise ratios of ($A, E,
T$) as well as the optimal signal-to-noise ratio.  For an assumed $h =
10^{-23}$, the SNRs of the three modes are plotted versus frequency.
For the equal-arm case computed here, the SNRs of $A$ and $E$ are
equal across the band.  In the long wavelength region of the band,
modes $A$ and $E$ have SNRs much greater than mode $T$, where its
contribution to the total SNR is negligible. At higher frequencies,
however, the $T$ combination has SNR greater than or comparable to the
other modes and can dominate the SNR improvement at selected
frequencies. Some of these results have also been obtained in \cite{NPDV03_1}.

\subsection{Optimization of SNR for binaries with known direction but
  with unknown orientation of the orbital plane}

Binaries will be important sources for LISA and therefore the analysis
of such sources is of major importance. One such class is of massive
or super massive binaries whose individual masses could range from
$10^3 M_{\odot}$ to $10^8 ~ M_{\odot}$ and which could be upto a few
Gpc away. Another class of interest are known binaries within our own
galaxy whose individual masses are of the order of a solar mass but
are just at a distance of a few kpc or less. Here the focus will be on
this latter class of binaries. It is assumed that the direction of the
source is known, which is so for known binaries in our galaxy.
However, even for such binaries, the inclination angle of the plane of
the orbit of the binary is either poorly estimated or unknown. The
optimization problem is now posed differently: the SNR is optimized
{\it after} averaging over the polarizations of the binary signals, so
the results obtained are optimal on the average, that is, the source
is tracked with an observable which is optimal on the average
\cite{NPDV03_1}. For computing the average, a uniform distribution for
the direction of the orbital angular momentum of the binary is
assumed.
 
When the binary masses are of the order of a solar mass and the
signal typically has a frequency of a few mHz, the GW frequency of
the binary may be taken to be constant over the period of
observation, which is typically taken to be of the order of an year.
A complete calculation of the signal matrix and the optimization
procedure of SNR, is given in \cite{NDPV03_2}. Here we briefly
mention the main points and the final results.

\begin{figure}
\begin{center}
\caption{Apparent position of the source in the sky as seen from LISA
  frame for \break $(\theta_B= 90^{\circ}, \,\phi_B=0^{\circ} \,)$.
  The track of the source for a period of one year is shown on the
  unit sphere in the LISA reference frame.} 
\includegraphics{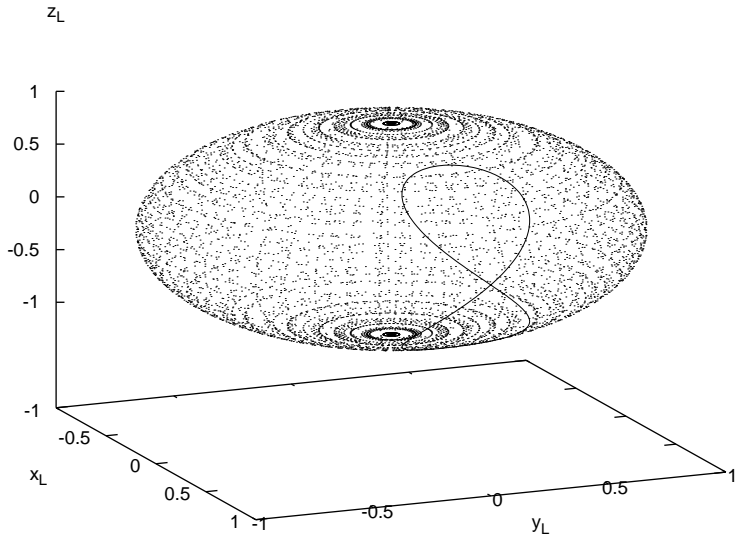}
\label{trk}
\end{center}
\end{figure}

A source fixed in the Solar System Barycentric reference frame in the
direction $(\theta_B, \phi_B)$ is considered. But as the LISA
constellation moves along its heliocentric orbit, the apparent
direction $(\theta_L, \phi_L)$ of the source in the LISA reference
frame $(x_L, y_L, z_L)$ changes with time. The LISA reference frame
$(x_L, y_L, z_L)$ has been defined in reference \cite{NDPV03_2} as
follows: the origin lies at the center of the LISA triangle and the
plane of LISA coincides with the $(x_L, y_L)$ plane with spacecraft 2
lying on the $x_L$ axis.  Fig. (\ref{trk}) displays this apparent
motion for a source lying in the ecliptic plane, that is with
$\theta_B = 90^{\circ}$ and $\phi_B = 0^{\circ}$. The source in the
LISA reference frame describes a figure of 8.  Optimizing the SNR amounts to
tracking the source with an optimal observable as the source
apparently moves in the LISA reference frame .

Since an average has been taken over the orientation of the orbital
plane of the binary or equivalently over the polarizations, the signal
matrix ${\bf A}$ is now of rank 2 instead of rank 1 as compared with
the application in the previous subsection. The mutually orthogonal
data combinations $A, E, T$ are convenient in carrying out the
computations because in this case as well, they simultaneously
diagonalize the signal and the noise covariance matrix. The
optimization problem now reduces to an eigenvalue problem with the
eigenvalues being the squares of the SNRs. There are two eigen-vectors
which are labelled as $\vec{v}_{+, \times}$ belonging to two non-zero
eigenvalues. The two SNRs are labelled as, ${\rm SNR}_+$ and ${\rm
  SNR}_{\times}$ corresponding to the two orthogonal (thus
statistically independent) eigen-vectors $\vec{v}_{+, \times}$. As was
done in the previous subsection the two SNRs can be squared and added to
yield a network SNR, which is defined through the equation:

\be 
{\rm SNR}_{\rm network}^2 =  {\rm SNR}_+^2 + {\rm SNR}_{\times}^2 \ .
\ee
The corresponding observable is called as the network observable. The
third eigenvalue is zero and the corresponding eigenvector orthogonal
to $\vec {v}_{+}$ and $\vec{v}_{\times}$ gives zero signal.

\begin{figure}
\begin{center}
\caption{Sensitivity curves for the observables: 
Michelson, $\max[X, Y, Z]$, $\vec{v}_+$ and 
network for the source direction ($\theta_B = 90^{\circ}, \phi_B = 0^{\circ}$).}
\includegraphics{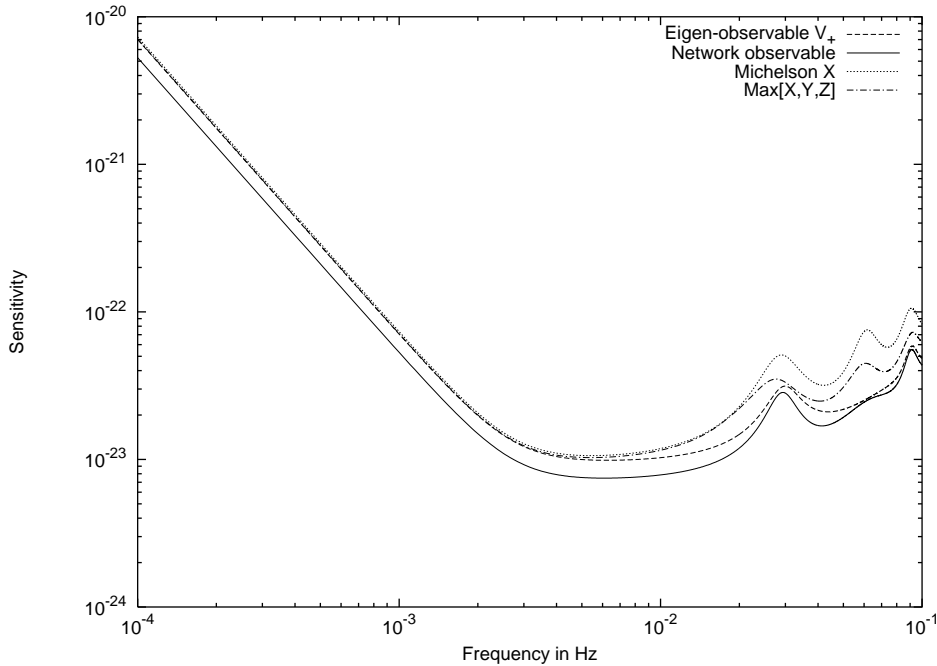}
\label{fig:sens}
\end{center}
\end{figure}

The eigenvectors and the SNRs are functions of the apparent source
direction parameters $(\theta_L, \phi_L)$ in the LISA reference frame, which in turn are
functions of time. The eigenvectors optimally track the source as it
moves in the LISA reference frame. Assuming an observation period of an year,
the SNRs are integrated over this period of time. The sensitivities
are computed according to the procedure described in the previous
subsection \ref{GA}. The results of these findings are displayed in
Fig.\ref{fig:sens}.
It shows the sensitivity curves of the following observables:

\begin{enumerate}
\item The Michelson combination  $X$ (faint solid curve).  
\item The observable obtained by taking the maximum sensitivity among
  $X, Y$ and $Z$ for each direction, where $Y$ and $Z$ are the
  Michelson observables corresponding to the remaining two pairs of
  arms of LISA \cite{AET99}. This maximum is denoted by $\max [X, Y,
  Z]$ (dash-dotted curve) and is operationally given by switching the
  combinations $X, Y, Z$ so that the best sensitivity is achieved.
\item The eigen-combination $\vec{v}_+$ which has the best sensitivity
  among all data combinations (dashed curve).
\item The network observable (solid curve).
\end{enumerate}

\begin{figure}
\caption{Ratios of the sensitivities of the observables network, $\vec{v}_{+, \times}$ with 
$\max [X, Y, Z]$ for the source direction $\theta_B = 90^{\circ}, \phi_B = 0^{\circ}$.}
\includegraphics[width=0.5\textwidth]{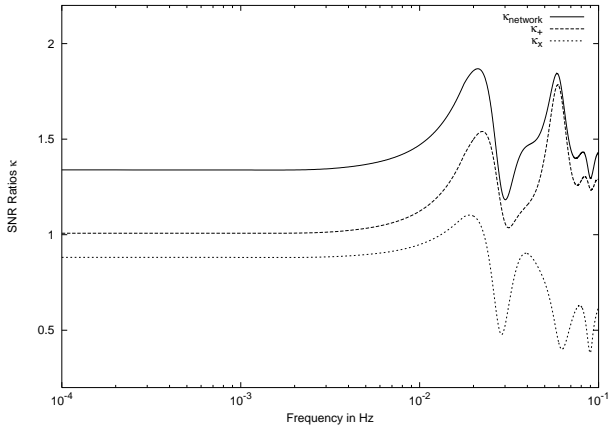}
\label{fig:SNRratio}
\end{figure}

It is observed that the sensitivity over the band-width of LISA increases
as one goes from (1) to (4). Also it is seen that the $\max [X, Y, Z]$ does
not do much better than $X$. This is because for the source direction
chosen $\theta_B = 90^{\circ}$, $X$ is reasonably well oriented and
switching to $Y$ and $Z$ combinations does not improve the sensitivity
significantly. However, the network and $\vec{v}_+$ observables show
significant improvement in sensitivity over both $X$ and $\max [X, Y,
Z]$. This is the typical behavior and the sensitivity curves (except
$X$) do not show much variations for other source directions and the
plots are similar. Also it may be fair to compare the optimal
sensitivities with $\max [X, Y, Z]$ rather than $X$. This comparison
of sensitivities is shown in Fig.\ref{fig:SNRratio}, where the network and 
the eigen-combinations $\vec{v}_{+, \times}$ are compared with $\max [X, Y, Z]$.

 Defining:
\begin{equation}   
\kappa_a (f) = \frac{{\rm SNR}_a(f)}{{\rm SNR}_{\max [X, Y, Z]}(f)} \, ,
\end{equation}
where the subscript $a$ stands for network or $+, \times$ and ${\rm
  SNR}_{\max [X, Y, Z]}$ is the SNR of the observable $\max [X, Y,
Z]$, the ratios of sensitivities are plotted over the LISA band-width.
The improvement in sensitivity for the network observable
is about 34$\%$ at low frequencies and rises to nearly 90$\%$ at about
20 mHz, while at the same time the $\vec{v}_+$ combination shows
improvement of 12$\%$ at low frequencies rising to over 50$\%$ at
about 20 mHz.

\section{Concluding Remarks}
\label{SECVII}

In this article we have summarized the use of TDI for canceling the
laser phase noise from heterodyne phase measurements performed by a
constellation of three spacecraft tracking each other along arms of
unequal length.  Underlying the TDI technique is the mathematical
structure of the theory of Gr\"obner basis and the algebra of modules
over polynomial rings. These methods have been motivated and
illustrated with the simple example of an unequal arm interferometer
in order to give a physical insight of TDI. Here, these methods have
been rigorously applied to the idealized case of a stationary LISA for
deriving the generators of the module from which the entire TDI data
set can be obtained; they can be extended in a straightforward way to
more than three spacecraft for possible LISA follow-on missions. The
stationary LISA case was used as a propaedeutical introduction to the
physical motivation of TDI, and for further extending it to the
realistic LISA configuration of free-falling spacecraft orbiting
around the Sun. The TDI data combinations canceling laser phase noise
in this general case are referred to as {\it second generation TDI},
and they contain twice as many terms as their corresponding {\it first
  generation} combinations valid for the stationary configuration.
\par
As a data analysis application we have shown that it is possible to
identify specific TDI combinations that will allow LISA to achieve
optimal sensitivity to gravitational radiation
\cite{PTLA02,NPDV03_1,NDPV03_2}. The resulting improvement in
sensitivity over that of an unequal-arm Michelson Interferometer, in
the case of monochromatic signals randomly distributed over the
celestial sphere and of random polarization, is non negligible. We
have found this to be equal to a factor of $\sqrt{2}$ in the low-part
of the frequency band, and slightly more than $\sqrt{3}$ in the
high-part of the LISA band.  The SNR for binaries whose location in
the sky is known but their polarization is not can also be optimized,
and the degree of improvement depends on the location of the source in
the sky.
\par
As a final remark we would like to emphasize that this field of
research, TDI, is still very young and evolving. Possible physical
phenomena, yet unrecognized, might turn out to be important to account
for within the TDI framework. The purpose of this review was to
provide the basic mathematical tools needed for working on future TDI
projects. We hope to have accomplished this goal, and that others will
be stimulated to work in this new and fascinating field of research.

\section*{Acknowledgement}
\noindent
SVD acknowledges support from IFCPAR, Delhi, India under which the
work was carried out in collaboration with J-Y Vinet. This research
was performed at the Jet Propulsion Laboratory, California Institute
of Technology, under contract with the National Aeronautics and Space
Administration.

\appendix
\section{Generators of the Module of Syzygies}
\label{syz}

We require the 4-tuple solutions $(q_3, q_1', q_2', q_3')$ to the equation:
\be
(1 - xyz) q_3 + (xz - y) q_1' + x(1 - z^2) q_2' + (1 - x^2) q_3' = 0 \, ,
\label{cnstr1}
\ee 
where for convenience we have substituted $x = \D_1, y = \D_2, z =
\D_3$. $q_3, q_1', q_2', q_3'$ are polynomials in $x, y, z$ with
integral coefficients i.e. in Z[x,y,z].

We now follow the procedure in the book by Becker et al.~\cite{becker}.

Consider the ideal in $Z[x,y,z]$ (or ${\cal Q}[x,y,z]$ where ${\cal
  Q}$ denotes the field of rational numbers), formed by taking linear
combinations of the coefficients in Eq.(\ref{cnstr1}) $f_1 = 1 - xyz,
f_2 = xz - y, f_3 = x(1 - z^2), f_4 = 1 - x^2$.  A Gr\"obner basis for
this ideal is: \be {\cal G} = \{g_1 = z^2 - 1, g_2 = y^2 - 1, g_3 = x
- yz \} \, .  \ee The above Gr\"obner basis is obtained using the
function GroebnerBasis in Mathematica.  One can check that both the
$f_i, i = 1,2,3,4$ and $g_j, j =1,2,3$ generate the same ideal because
we can express one generating set in terms of the other and
vice-versa: \be f_i = d_{ij} g_j , ~~~~~~~~g_j = c_{ji} f_i \, , \ee
where $d$ and $c$ are $4 \times 3$ and $3 \times 4$ polynomial
matrices respectively, and are given by, \be d =
\left(\begin{array}{ccc}
    -1 & -z^2 & -yz \\
    y & 0 & z \\
    -x & 0 & 0 \\
    -1 & - z^2 & -(x + yz)
\end{array}\right) \, ,
 \hspace{0.3in}
c = \left(\begin{array}{cccc}
  0 & 0 & -x & z^2 - 1 \\
  -1 & -y & 0 & 0 \\
  0 & z & 1 & 0 
\end{array}\right) 
\,.
\label{cd}
\ee 
The generators of the 4-tuple module are given by the set $A \,
\bigcup B^*$ where $A$ and $B^*$ are the sets described below:

$A$ is the set of row vectors of the matrix $I - d \cdot c $ where the
dot denotes the matrix product and I is the identity matrix, $4 \times
4$ in our case. Thus, \bea
a_1 &=& (z^2 - 1, 0, x - yz, 1 - z^2) \, , \no \\
a_2 &=& (0, z(1 - z^2), xy - z, y(1 - z^2)) \, , \no \\
a_3 &=& (0, 0, 1 - x^2, x(z^2 - 1)) \, , \no \\
a_4 &=& (-z^2, xz, yz, z^2) \, .  
\eea 
We thus first get 4 generators.  The additional generators are
obtained by computing the S-polynomials of the Gr\"obner basis ${\cal
  G}$.  The S-polynomial of two polynomials $g_1, g_2$ is obtained by
multiplying $g_1$ and $g_2$ by suitable terms and then adding, so that
the highest terms cancel. For example in our case $g_1 = z^2 - 1$ and
$g_2 = y^2 - 1$ and the highest terms are $z^2$ for $g_1$ and $y^2$
for $g_2$ . Multiply $g_1$ by $y^2$ and $g_2$ by $z^2$ and subtract.
Thus, the S-polynomial $p_{12}$ of $g_1$ and $g_2$ is: \be p_{12} =
y^2 g_1 - z^2 g_2 = z^2 - y^2 \, .  \ee Note that order is defined ($x
>> y >> z$) and the $y^2 z^2$ term cancels.  For the Gr\"obner basis
of 3 elements we get 3 S-polynomials $p_{12}, p_{13}, p_{23}$.  The
$p_{ij}$ must now be re-expressed in terms of the Gr\"obner basis ${\cal
  G}$.  This gives a $3 \times 3$ matrix $b$. The final step is to
transform to 4-tuples by multiplying $b$ by the matrix $c$ to obtain
$b^* = b \cdot c$. The row vectors $b^*_i , i = 1, 2, 3 $ of $b^*$
form the set $B^*$: 
\bea
b^*_1 &=& (z^2 - 1, y(z^2 - 1), x(1 - y^2), (y^2 -1)(z^2 - 1)) \, , \no \\
b^*_2 &=& (0, z(1 - z^2), 1 - z^2 - x(x - yz), (x - yz)(z^2 - 1)) \, , \no \\
b^*_3 &=& (-x + yz, z - xy, 1 - y^2, 0) .  
\eea 
Thus we obtain 3 more generators which gives us a total of 7
generators of the required module of syzygies.

\section{Conversion between generating sets}
\label{mat}

We list the three sets of generators and relations among them. We
first list below $\a, \b, \g, \z$: \bea
\a &=& (-1, -z, -xz, 1, xy, y) \, , \no \\
\b &=& (-xy, -1, -x, z, 1, yz) \, , \no \\
\g &=& (-y, -yz, -1, xz, x, 1) \, , \no \\
\z &=& (-x, -y, -z, x, y, z) \, .  \eea

We now express the $a_i$ and $b^*_j$ in terms of $\a, \b, \g, \z$: 
\bea
a_1 &=& \g - z \z \, , \no \\
a_2 &=& \a - z \b \, , \no \\
a_3 &=& - z \a + \b - x \g + xz \z \, , \no \\
a_4 &=& z \z \no \\
b^*_1 &=& - y \a + yz \b + \g - z \z \, , \no \\
b^*_2 &=& (1 - z^2) \b - x \g + xz \z \, , \no \\
b^*_3 &=& \b - y \z \, . 
\eea
Further we also list below $\a, \b, \g, \z$ in terms of $X^{(A)}$:
\bea
\a &=& X^{(3)} \, , \no \\
\b &=& X^{(4)} \, , \no \\
\g &=& - X^{(1)} + z X^{(2)} \, , \no \\
\z &=& X^{(2)} \, . 
\eea
This proves that since the $a_i, b^*_j$ generate the required module, the 
$\a, \b, \g, \z$ and \break $X^{(A)}, A = 1, 2, 3, 4$ also generate the same module.
\par
The Gr\"obner basis is given in terms of the above generators as follows: \break
$G^{(1)} = \z, G^{(2)}=  X^{(1)}, G^{(3)} = \b, G^{(4)} = \a$ and
$G^{(5)} = a_3$


\begin{references}

\bibitem{PBDH} P.L. Bender and D. Hils, \emph{CQG}, \textbf{14}, 1439(1997).

\bibitem{NYPZ} G. Nelemans, L.R. Yungelson and S.F. Portegies Zwart, 
  \emph{A \& A}, \textbf{375}, 890(2001).
  
\bibitem{TE95} M. Tinto, and F.B. Estabrook, {\it Phys. Rev. D}, {\bf 52},
1749, (1995).

\bibitem{Cass} M. Tinto, {\it Class. Quantum Grav.}, {\bf 19}, 7, 1767 (2002).

\bibitem{EW75} F.B. Estabrook, and H.D. Wahlquist, {\it Gen. Relativ. Gravit.}
{\bf 6}, 439 (1975).

\bibitem{T02} M. Tinto, {\it Class. Quantum Grav.}, {\bf 19},  1767 (2002)
  
\bibitem{LISA98} LISA: (Laser Interferometer Space Antenna) {\it An
    international project in the field of Fundamental Physics in
    Space}, Pre-Phase A Report, {\bf MPQ 233}, (Max-Planck-Institute
  f\"ur Quantenoptic, Garching bei M\"unchen, 1998).
  
\bibitem{TA99} M. Tinto, and J.W. Armstrong {\it Phys. Rev. D}, {\bf
    59}, 102003 (1999).
  
\bibitem{TEA02} M. Tinto, F.B.  Estabrook, and J.W. Armstrong, {\it
    Phys. Rev. D}, {\bf 65}, 082003 (2002).

\bibitem{T98} M. Tinto, {\it Phys. Rev. D}, {\bf 58}, 102001, (1998).
  
\bibitem{AET99} J.W. Armstrong, and F.B. Estabrook, \& M. Tinto, {\it
    ApJ}, {\bf 527}, 814 (1999).
  
\bibitem{DNV02} S.V. Dhurandhar, K. R. Nayak, and J.-Y. Vinet {\it
    Phys. Rev. D}, {\bf 65}, 102002 (2002)
  
\bibitem{ETA00} F.B. Estabrook, M. Tinto, and J.W. Armstrong, {\it
    Phys. Rev. D}, {\bf 62}, 042002 (2000).

\bibitem{S03} D.A. Shaddock, {\it Phys. Rev. D}, {\bf 69 }, 022001  (2004).

\bibitem{CH03} N.J. Cornish \& R.W. Hellings, {\it Class. Quantum Grav.}, {\bf 20}, 4851, (2003).
  
\bibitem{STEA03} D.A. Shaddock, M. Tinto, F.B. Estabrook, and J.W.
  Armstrong, {\it Phys. Rev. D}, {\bf 68}, 061303 (2003).
  
\bibitem{TEA04} M. Tinto, F.B.  Estabrook, and J.W. Armstrong, {\it
    Phys. Rev. D}, {\bf 69}, 082001, (2004).
  
\bibitem{PTLA02} T. A. Prince, M. Tinto, S. L. Larson, and J.W. Armstrong, 
    {\it Phys.Rev. D} {\bf 66}, 122002 (2002).

\bibitem{NPDV03_1} K. Rajesh Nayak, A. Pai, S. V. Dhurandhar, and
  J-Y. Vinet, \textit{Class. Quantum Grav}, \textbf{20,} 1217(2003).
  
\bibitem{NDPV03_2} K. Rajesh Nayak, S. V. Dhurandhar, A. Pai, and J-Y.
  Vinet, {\it Phys. Rev. D}, {\bf 68}, 122001 (2003).
  
\bibitem{FB84} J.E. Faller and P.L. Bender, in: {\it Precision
    Measurement and Fundamental Constants II}, eds. B.N. Taylor and
  W.D. Phillips, NBS Spec. Pub. 617, (1984).

\bibitem{FBHHV85} J.E. Faller, P.L. Bender, J.L. Hall, D. Hils and M.A. Vincent, in:
{\it Proceedings Colloquium Kilometric Optical Arrays in Space}, ESA SP-126,
Noordwijk, The Netherlands (1985).

\bibitem{FBHHSV89} J.E. Faller, PL. Bender, J.L. Hall, D. Hils, R.T.
  Stebbins and M.A. Vincent, in: {\it Advances in Space Research},
  Proc. XXVII COSPAR, Symp. 15 on Relativistic Gravitation, Vol. 9
  (1989).

\bibitem{GHTF96} G. Giampieri, R.W. Hellings, M. Tinto, and
  J.E. Faller, {\it Optics Communications}, {\bf 123}, 669 (1996).
  
\bibitem{JW68} G. M. Jenkins and D. G. Watts, {\it Spectral Analysis and
    Its Applications}, (Holden-Day, San Francisco), (1968).

\bibitem{Wolf02} S. Wolfram, {\it Mathematica: User
    manual}, Wolfram Research, Inc., (2002).
  
\bibitem{becker} T.~Becker, {\it "Gr\"obner Bases: A computational 
approach to commutative algebra"}, Volker W

\bibitem{KR}
M.~Kreuzer and  L.~Robbiano, {\it "Computational commutative
 algebra 1"}, Springer Verlag(2000).

\bibitem{Summers} D. Summers, "Algorithm tradeoffs," oral
  presentation, 3rd progress meeting of the ESA funded LISA PMS
  Project. ESTEC, NL, February 2003.

\bibitem{Folkner} W.M. Folkner, F. Hechler, T.H. Sweetser, M.A. Vincent, and
  P.L. Bender, {\it Clas. Quantum Grav.}, {\bf 14}, 1543, (1997). 

\bibitem{TAE01} M. Tinto, J.W. Armstrong \& F.B. Estabrook, {\it Phys. Rev. D}, {\bf 63}, 021101 (2001)

\bibitem{NV04} R. Nayak and J-Y Vinet, (under preparation), (2004).

\bibitem{Noble69} B. Noble, {\it  Applied Linear Algebra},
  Prentice/Hall International, p. 378, (1969).
  
\bibitem{Selby64} S. Selby, {\it Standard of
  Mathematical Tables}, The Chemical Bubber Co., p. 131, (1964). 

\bibitem{Finn01} L.S. Finn, {\it Phys. Rev. D}, {\bf 63},
  102001, (2001).
  
\bibitem{WP02} M. Tinto, F.B.
  Estabrook, \& J.W. Armstrong, {\it LISA Pre-Project Publication},
  $(http://www.srl.caltech.edu/lisa/tdi\_wp/LISA\_Whitepaper.pdf)$,
  (2002).

\end{references}
\end{document}